\documentstyle[12pt]{article}
\input xy
\xyoption{all}

\begin{document}


\hfill $\,$

\vspace{1.0in}

\begin{center}

{\large\bf GLSM's for Partial Flag Manifolds}

\vspace{0.2in}

Ron Donagi$^1$, Eric Sharpe$^2$\\
$^1$ Department of Mathematics\\
University of Pennsylvania\\
David Rittenhouse Lab.\\
209 South 33rd St.\\
Philadelphia, PA  19104-6395\\
$^2$ Departments of Physics, Mathematics \\
University of Utah \\
Salt Lake City, UT  84112 \\
{\tt donagi@math.upenn.edu},
{\tt ersharpe@math.utah.edu} \\

$\,$

\end{center}

In this paper we outline some aspects of nonabelian gauged linear
sigma models.  First, we review how partial flag manifolds (generalizing
Grassmannians) are described physically by nonabelian gauged linear sigma
models, paying attention to realizations of tangent bundles and other
aspects pertinent to (0,2) models.  Second, we review constructions
of Calabi-Yau complete intersections within such flag manifolds, and
properties of the gauged linear sigma models.  We discuss a number
of examples of nonabelian GLSM's in which the K\"ahler phases
are not birational, and in which at least one phase is realized in
some fashion other than as a complete intersection, extending previous
work of Hori-Tong.  We also review an example of an abelian GLSM exhibiting
the same phenomenon.  We tentatively identify the mathematical
relationship between such non-birational phases, as examples of
Kuznetsov's homological projective duality.  Finally, we discuss
linear sigma model moduli spaces in these gauged linear sigma models.
We argue that the moduli spaces being realized physically by these
GLSM's are precisely Quot and hyperquot schemes, as one would expect
mathematically.

\begin{flushleft}
April 2007
\end{flushleft}

\newpage

\tableofcontents

\newpage

\section{Introduction}

Two-dimensional gauged linear sigma models \cite{WitPhases} have proven
to be an extremely powerful tool for describing a wide variety of spaces.
Their primary original use was for describing toric varieties and
complete intersections therein.  More recently \cite{stxglsm} they have
been used to describe toric stacks and complete intersections therein.

In addition to toric manifolds, they have also been used to describe
Grassmannians \cite{WitPhases,witver}, and although this application
was comparatively obscure, recently it has been explored in more
depth \cite{hori1}.  In this paper we shall push this direction further
by outlining basic aspects of gauged linear sigma models for
partial flag\footnote{In this paper, we shall use the terms
``partial flag manifold'' and ``flag manifold''
interchangeably.  In particular, ``flag manifold'' should not be
interpreted to mean only a complete flag manifold.} manifolds, 
which generalize Grassmannians.

We begin in section~\ref{flagmflds} 
with an overview of how flag manifolds are realized in physics
via nonabelian gauged linear sigma models, also describing various dualities
that exist among the flag manifolds, Pl\"ucker coordinates for
flag manfiolds, and more complicated manifolds and stacks realizable with
gauged linear sigma models.
In section~\ref{flagbdles} we describe how tangent bundles of flag manifolds
are realized physically in gauged linear sigma models,
a necessary step for considering (0,2) theories.

Next, in section~\ref{ciflag} 
we discuss Calabi-Yau complete intersections in flag manifolds.
As in \cite{hori1}, we also examine K\"ahler phases of some gauged linear
sigma models, and describe a number of examples in which the different
K\"ahler phases have geometric interpretations, sometimes realized
in novel fashions, but are not related
by birational transformations.  In the past, it was thought that geometric
phases of gauged linear sigma models must be related by birational 
transformations, but recently in \cite{ps4}[section 12.2] 
and \cite{hori1},
counterexamples have appeared.  We interpret the different phases as
being related by Kuznetsov's homological projective duality
\cite{kuz1,kuz2,kuz3}, an idea that will be explored much further in 
\cite{alltoappear}.  We use that proposal that gauged linear sigma models
realize Kuznetsov's duality to make a proposal for the physical interpretation
of Calabi-Yau complete intersections in $G(2,N)$ for $N$ even,
which has been mysterious previously.
We also briefly outline how such non-birational phase
phenomena can appear in abelian GLSM's, reviewing an example that first
appeared in \cite{ps4}[section 12.2] and which will be explored in greater 
detail (together with many other examples)
in \cite{alltoappear}.

Finally, in section~\ref{quantcohom}
we review how quantum cohomology of Grassmannians and flag
manifolds is realized in corresponding GLSM's, and describe 
linear sigma model moduli spaces in nonabelian GLSM's.
Physically, the computation of linear sigma model moduli spaces
for nonabelian GLSM's has not been clear, unlike the case of
abelian GLSM's.
Mathematically, on the other hand,
there are known candidates for what such spaces should be -- 
Quot and hyperquot schemes, specifically -- that have been used by
mathematicians when studying quantum cohomology rings of
flag manifolds.
In this paper, we show in examples that
LSM moduli spaces realized
by nonabelian GLSM's are precisely Quot and hyperquot schemes,
as one would have expected naively, instead of, for example,
some different spaces birationally equivalent to Quot and hyperquot schemes.

\section{Flag manifolds}   \label{flagmflds}

\subsection{Basics}

A flag manifold is a manifold describing possible configurations
of a series of nested vector spaces.
We will use the notation $F(k_1, k_2, \cdots, k_n, N)$ to
indicate configuration space
of all $k_1$-dimensional planes sitting inside $k_2$-dimensional
planes which themselves sit inside $k_3$-dimensional planes, and so
forth culminating in $k_n$-dimensional planes sitting inside the
vector space ${\bf C}^N$.

Grassmannians are a special case of flag manifolds.
The Grassmannian of $k$-planes inside ${\bf C}^N$, denoted
$G(k,N)$, is the same as the flag manifold $F(k,N)$.
The space $G(k,N)$ has complex dimension $k(N-k)$, and
Euler characteristic
\begin{displaymath}
\left( \begin{array}{c}
N \\ k \end{array} \right) \: = \:
\frac{ N! }{ k! (N-k)!}
\end{displaymath}
As one special case, $G(1,N)$ is the projective space ${\bf P}^{N-1}$,
so Grassmannians themselves generalize projective spaces.

Flag manifolds can be described as cosets as follows:
\begin{equation}    \label{flagcoset}
F(k_1, k_2, \cdots, k_n, N) \: = \: \frac{ U(N) }{
U(k_1) \times U(k_2-k_1) \times \cdots U(k_n-k_{n-1}) \times
U(N - k_n) }
\end{equation}
which has complex dimension
\begin{displaymath}
\sum_{i=1}^n \left( k_i \: - \: k_{i-1} \right)\left(
N \: - \: k_i \right) \: = \: \sum_{i=1}^n k_i\left( k_{i-1} \: - \: k_i
\right) \: + \: N k_n
\end{displaymath}
in conventions where $k_0 = 0$.
In particular, as a special case,
\begin{displaymath}
G(k,N) \: = \: \frac{ U(N) }{U(k) \times U(N-k)}
\end{displaymath}
which has complex dimension $k(N-k)$.

The Euler characteristic of $F(k_1, \cdots, k_n, N)$ is given
by the multinomial coefficient
\begin{displaymath}
\frac{ N! }{ k_1! (k_2-k_1)! \cdots (k_n - k_{n-1})! (N-k_n)! }
\end{displaymath}
Flag manifolds have only even-dimensional cohomology, with a basis given
by Schubert classes, which correspond to elements of $W/W_P$
(in a description of the flag manifold as $G/P$ for $P$ a parabolic
subgroup of $G$).  Thus,
$\chi(G/P) = |W|/|W_P|$.

There are natural projection maps to flag manifolds with fewer flags,
\begin{displaymath}
F(k_1, \cdots, k_n, N) \: \longrightarrow \: F(k_1, \cdots, 
\hat{k_i}, \cdots, k_n, N)
\end{displaymath}
(where $\hat{k_i}$ denotes an omitted entry)
defined by forgetting the intermediate flags.
The projection map above 
has fiber $G(k_i-k_{i-1}, k_{i+1}-k_{i-1})$.
For example, there is a map
\begin{displaymath}
F(1,2,n) \: \longrightarrow \: F(1,n) = G(1,n)
\end{displaymath}
with fiber $G(1,n-1)$.

These manifolds also have alternative descriptions which make it
possible to describe them with gauged linear sigma models.
The Grassmannian $G(k,N)$ can be described as
\begin{displaymath}
G(k,N) \: = \: {\bf C}^{kN} // GL(k)
\end{displaymath}
which corresponds physically to a two-dimensional $U(k)$
gauge theory with $N$ chiral fields in the fundamental of $U(k)$.
The D-terms in the supersymmetric gauge theory have the form
\cite{WitPhases}
\begin{displaymath}
D^i_j \: = \: \sum_{s=1}^N \phi^{is} \overline{\phi}_{js}
\: - \: r \delta^i_j
\end{displaymath}
We can interpret the sum above as the dot product of two vectors
in ${\bf C}^N$, and so for $r \gg 0$,  we see that the vectors
must be orthogonal and normalized.  In other words, the D-terms force
the $\phi$ fields to describe a $k$-dimensional subspace of ${\bf C}^N$,
exactly right to describe the Grassmannian of $k$ planes in ${\bf C}^N$.

Similarly, flag manifolds can be described as
\begin{eqnarray*}
\lefteqn{ F(k_1, k_2, \cdots, k_n, N) \: = \: } \\
&& \left(  {\bf C}^{k_1 k_2} \times {\bf C}^{k_2 k_3} \times
{\bf C}^{k_3 k_4} \times \cdots \times {\bf C}^{k_{n-1} k_n} \times
{\bf C}^{k_n N} \right) // \left( 
GL(k_1) \times GL(k_2) \times \cdots \times GL(k_n) \right)
\end{eqnarray*}
which corresponds physically to a two-dimensional
\begin{displaymath}
U(k_1) \times U(k_2) \times \cdots \times U(k_n)
\end{displaymath}
gauge theory with bifundamentals.

The D-term analysis is similar to that for Grassmannians.
The D-terms for the first $U(k_1)$ factor are of the form
\begin{displaymath}
D^i_j \: = \: \sum_{s=1}^{k_2} \phi^{is} \overline{\phi}_{js}
\: - \: r_1 \delta^i_j
\end{displaymath}
identical to those for Grassmannians, and for $r_1 \gg 0$ the conclusion
is the same:  the $\phi$'s define $k_1$ orthogonal, normalized vectors
in ${\bf C}^{k_2}$.
A slightly more interesting case is a generic $U(k_i)$ factor.
Here the D-terms are of the form
\begin{displaymath}
D^i_j \: = \: - \sum_{s=1}^{k_{i-1}} \phi^{is} \overline{\phi}_{js}
\: + \: \sum_{a=1}^{k_{i+1}} \tilde{\phi}^{ai} \overline{\tilde{\phi}}_{aj}
\: - \: r_i \delta^i_j
\end{displaymath}
where the $\phi$'s are $({\bf k_{i-1}}, {\bf \overline{k_i}})$
bifundamentals and the $\tilde{\phi}$'s are $({\bf k_i}, 
{\bf \overline{k_{i+1}}})$
bifundamentals.
If we make the inductive assumption that the $\phi$'s define
$k_{i-1}$ orthogonal, normalized vectors in ${\bf C}^{k_i}$,
then the first sum above is proportional to $r_{i-1} \delta^i_j$
for some $i$, $j$, and vanishes for others.  So long as both $r_{i-1} \gg 0$
and $r_i \gg 0$, the result is the same:  the D-terms give a constraint on
the $\tilde{\phi}$, which implies that they are orthogonal, normalized
vectors in ${\bf C}^{k_{i+1}}$.  Proceeding inductively we see that so
long as all the Fayet-Iliopoulos parameters are positive, we recover
the flag manifold.

In this language, we can define the (forgetful) projection map
\begin{displaymath}
F(k_1, k_2, \cdots, k_n, N) \: \longrightarrow \: F(k_1, \cdots, 
\hat{k_i}, \cdots, k_n, N)
\end{displaymath}
(where $\hat{k_i}$ denotes an omitted entry) as follows.
Let $B^a_b$ denote the bifundamentals transforming in the
$({\bf k_{i-1}}, {\bf \overline{k_i}})$ representation of $U(k_{i-1})
\times U(k_i)$, and $C^b_c$ denote the bifundamentals transforming
in the $({\bf k_i}, {\bf \overline{k_{i+1}}})$ representation of
$U(k_i) \times U(k_{i+1})$, then the projection map is defined by
removing the $U(k_i)$ projection factor and replacing $B$, $C$
with the factor $BC$ where the internal $U(k_i)$ index is summed over.

For more information on flag manifolds and Grassmannians,
see for example \cite{manin,hubsch}.

\subsection{Pl\"ucker coordinates and baryons}

Grassmannians and flag manifolds can be embedded into projective
spaces using ``Pl\"ucker coordinates,'' which can be understood
physically as baryons in the gauged linear sigma model.
Let us first review how this works for Grassmann manifolds, then
describe the construction for flag manifolds.

Describe a Grassmannian $G(k,N)$ as a $U(k)$ gauge theory with
$N$ chiral fields $A^i_s$ 
transforming in the fundamental representation of $U(k)$.
The $SU(k)$ invariant field configurations (``baryons,'' for mathematical 
readers) have the form
\begin{displaymath}
P_{s_1 \cdots s_k} \: = \:
\epsilon_{i_1 \cdots i_k} A_{s_1}^{i_1} A_{s_2}^{i_2} \cdots
A_{s_k}^{i_k}
\end{displaymath}
where $i$ is a $U(k)$ index and $s \in \{ 1, \cdots, N\}$.
This gives us
\begin{displaymath}
\left( \begin{array}{c}
N \\ k \end{array} \right) \: = \: \frac{N!}{k! (N-k)!}
\end{displaymath}
$SU(k)$-invariant composite fields.
Since the gauge group is $U(k)$ not $SU(k)$, each of these
composite fields transforms with the same weight under the overall $U(1)$,
and the zero locus is disallowed by D-terms, we interpret each of
these baryons as homogeneous coordinates on a projective space of
dimension
\begin{displaymath}
\left( \begin{array}{c}
N \\ k \end{array} \right) \: - \: 1
\end{displaymath}

We can follow a similar procedure to construct Pl\"ucker coordinates
for flag manifolds.
Consider the flag manifold constructed as the classical Higgs moduli space
of a
\begin{displaymath}
U(k_1) \times U(k_2) \times \cdots \times U(k_n)
\end{displaymath}
gauge theory with bifundamentals in the
$({\bf k_1}, {\bf \overline{k_2}})$ representation of $U(k_1) \times U(k_2)$,
$({\bf k_2}, {\bf \overline{k_3}})$ representation of $U(k_2) \times U(k_3)$,
and so forth, concluding with a bifundamental in the
$({\bf k_n}, {\bf \overline{N}})$ representation of $U(k_n) \times U(N)$.

Let $A^a_b$ denote the bifundamental of $U(k_1) \times U(k_2)$,
$B^b_c$ denote the bifundamental of $U(k_2) \times U(k_3)$,
and $C^c_d$ denote the bifundamental of $U(k_n) \times U(N)$.
If we contract all internal indices, then the product
\begin{displaymath}
A B \cdots C
\end{displaymath}
transforms as the $({\bf k_1}, {\bf \overline{N}})$ of
$U(k_1) \times U(N)$, and by taking determinants of $k_1 \times k_1$
submatrices, just as for the Grassmannian $G(k_1,N)$, we can build
one set of baryons, which we interpret as homogeneous coordinates
on a projective space of dimension
\begin{displaymath}
\left( \begin{array}{c}
N \\ k_1 \end{array} \right) \: - \: 1.
\end{displaymath}
Contracting internal indices, the product
\begin{displaymath}
B \cdots C
\end{displaymath}
transforms as the $({\bf k_2}, {\bf \overline{N}})$ representation of
$U(k_2) \times U(N)$, and by taking determinants of $k_2 \times k_2$
submatrices, just as for the Grassmannian $G(k_2,N)$, we can build
another set of baryons, which we interpret as homogeneous coordinates
on a projective space of dimension
\begin{displaymath}
\left( \begin{array}{c}
N \\ k_2 \end{array} \right) \: - \: 1
\end{displaymath}
We can continue this process, which at the end will construct
the last set of baryons as determinants of $k_n \times k_n$
submatrices of $C$.

More formally, each product of bifundamentals is defining a
(forgetful) map
\begin{displaymath}
F(k_1, k_2, \cdots, k_n, N) \: \longrightarrow \: G(k_i, N)
\end{displaymath}
for some $i$.
Constructing baryons from a product of bifundamentals
is building Pl\"ucker coordinates on the Grassmannians, and so
ultimately mapping the flag manifold into a product of projective
spaces:
\begin{displaymath}
F(k_1, \cdots, k_n, N) \: \longrightarrow \:
\prod_{i=1}^n {\bf P}^{N_i - 1}
\end{displaymath}
where
\begin{displaymath}
N_i \: = \: \left( \begin{array}{c} N \\ k_i \end{array} \right)
\end{displaymath}

\subsection{Presentation dependence}

In gauged sigma models, linear or nonlinear, there is often
a question of, or assumptions made concerning,
presentation-dependence, which is believed to be resolved
by renormalization group flow.
For example, the ${\bf P}^n$ model can be alternately presented
as either an ungauged nonlinear sigma model on ${\bf P}^n$,
or, a gauged linear sigma model with $n+1$ matter fields all of
charge $1$.  Mathematically, these are describing the same thing,
but physically the two theories are different.
In this particular case, it has been checked by a variety
of means that these two presentations are in the same universality class. 

Such presentation-dependence assumptions in principle often appear
in gauged linear sigma models:  for example, the supersymmetric
${\bf P}^1$ model and a model describing a degree-two hypersurface
in ${\bf P}^2$ are describing the same target space geometry,
and so had better be in the same universality class, though we are
not aware of any work specifically checking this.

In more complicated cases, presentation-dependence is a more
important issue.  For example, strings on stacks 
\cite{stxglsm,ps4,ps1,ps2,ps5}
are described concretely by universality classes
of gauged nonlinear sigma models, where the gauge group need be
neither finite nor effectively-acting.  Because several
naive consistency checks fail, a great deal of effort was expended
in especially \cite{stxglsm,ps2} to verify that universality classes
are independent of presentation.
Also, presentation-dependence issues arise in the application
of derived categories to physics \cite{ps5}, though there 
the presentations involve open strings, not gauged sigma models.

Here let us list some interesting examples of flag manifolds
which can be presented both with nonabelian gauge theories
and, alternately, with abelian gauge theories.

One example, discussed in \cite{hori1}, is the Grassmannian
$G(2,4)$ of 2-planes in ${\bf C}^4$, which can be understood
as a hypersurface in ${\bf P}^5$, as follows:
let $\phi_{ij} = - \phi_{ji}$, $i \in \{1, \cdots, 4\}$,
denote homogeneous coordinates on ${\bf P}^5$, then 
$G(2,4)$ is given by the hypersurface
\begin{displaymath}
\phi_{12} \phi_{34} \: - \: \phi_{13} \phi_{24} \: + \:
\phi_{14} \phi_{23}  \: = \: 0
\end{displaymath}
In fact, this is just the Pl\"ucker embedding:  each $\phi_{ij}$ is
a baryon on $G(2,4)$, and it is straightforward to check that the
equation above follows from the definition of Pl\"ucker coordinates.
Furthermore, this result is invariant under duality transformations.

Another example is the flag manifold $F(1,n-1,n)$,
which is the degree $(1,1)$ hypersurface in 
${\bf P}^{n-1} \times {\bf P}^{n-1}$.
This also can be understood as a Pl\"ucker embedding.
Let $A^1_a$, $B^a_i$, $a \in \{ 1, \cdots, n-1\}$, $i \in
\{ 1, \cdots, n\}$ denote the bifundamentals of the two-dimensional
gauge theory, then define Pl\"ucker coordinates
\begin{eqnarray*}
p(AB)_{i_1} & = & A^1_a B^a_{i_1} \\
p(B)_{i_1 \cdots i_{n-1}} & = & B^{a_1}_{i_1} \cdots B^{a_{n-1}}_{i_{n-1}}
\epsilon_{a_1 \cdots a_{n-1}}
\end{eqnarray*}
Define $q_i$'s to be the dual of the $p(B)$'s:
\begin{displaymath}
q_i \: = \: \frac{1}{(n-1)!} \epsilon_{i i_1 \cdots i_{n-1}}
p(B)_{i_1 \cdots i_{n-1}}
\end{displaymath}
then the image of the Pl\"ucker embedding in ${\bf P}^{n-1}
\times {\bf P}^{n-1}$ is the same as the degree $(1,1)$ hypersurface
in ${\bf P}^{n-1} \times {\bf P}^{n-1}$ given by
\begin{displaymath}
\sum_i p(AB)_i q_i \: = \: 0
\end{displaymath}
This follows from the identity that
\begin{displaymath}
\sum_i A^1_a B^a_i B^{a_1}_{i_1} \cdots B^{a_{n-1}}_{i_{n-1}}
\epsilon_{i i_1 \cdots i_{n-1}} \epsilon_{a_1 \cdots a_{n-1}}
\: = \: 0
\end{displaymath}

The reader should be careful to note that these examples are the
exception, not the rule:  in general, although Pl\"ucker coordinates
will define an embedding into a product of projective spaces,
typically that embedding cannot be described as a complete intersection
of hypersurfaces in those projective spaces.  For the embedding to
be given globally as a complete intersection, as in the examples
above, is rare.

\subsection{Duality}

The Grassmannian of $k$-planes in ${\bf C}^N$ is the same as the
Grassmannian of $N-k$ planes in ${\bf C}^N$:
\begin{displaymath}
G(k,N) \: \cong \: G(N-k,N)
\end{displaymath}
Physically, this is a duality between a two-dimensional $U(k)$
field theory with $N$ fundamentals and a two-dimensional $U(N-k)$
field theory with $N$ fundamentals.  
Since both gauge theories describe the same manifold,
following the usual procedure for linear sigma models one assumes
that the chiral rings of each theory match.
This is very reminiscent of
Seiberg duality in four dimensions \cite{seibergdual},
which relates $SU(k)$ gauge theories with $N$ fundamentals and
$SU(N-k)$ gauge theories with $N$ fundamentals, for which much of the
original justification came from comparing chiral rings.

Let us briefly consider how Pl\"ucker coordinates behave under
this duality.
If we let $p_{i_1, \cdots, i_{k}}$ denote Pl\"ucker coordinates
on $G(k,N)$ (where $p$ is antisymmmetric in its indices and all
take values in $\{ 1, \cdots, N\}$), and
$q_{i_1, \cdots, i_{N-k}}$ denote Pl\"ucker coordinates on $G(N-k,N)$,
then it can be shown \cite{hp}[chapter II.VII.3, theorem 1]
that they are related by
\begin{displaymath}
p_{i_1, \cdots, i_{k}} \: = \: \frac{1}{(N-k)!} \epsilon_{i_1, \cdots, i_N} \,
q_{i_{k+1}, \cdots, i_N}
\end{displaymath}
This will be useful when relating Calabi-Yau complete intersections
and bundles defined over dual Grassmannians.

For flag manifolds, there is an analogous duality:
the manifold of $k_1$-planes in $k_2$-planes in $k_3$-planes
and so forth is diffeomorphic to the manifold of $k_1$-planes
in $(k_3-k_2+k_1)$-planes in $k_3$ planes:
\begin{displaymath}
F(k_1, k_2, k_3, \cdots, N) \: \cong \:
F(k_1, k_3-k_2+k_1, k_3, \cdots, N)
\end{displaymath}
More generally, any entry $k_i$ can be replaced by
$k_{i+1}-k_i+k_{i-1}$, with the exception of $k_1$ which dualizes
to $k_2-k_1$.
It is straightforward to check that the coset 
representation~(\ref{flagcoset})
is symmetric under this duality.
However, although this duality relates flag manifolds by diffeomorphisms,
the resulting flag manifolds need not be biholomorphic. 
In general, this duality can be described explicitly using the metric.
A flag of dimensions $k_1, \cdots, k_n$ is equivalent to a collection
of mutually orthogonal subspaces of dimensions $k_1, k_2-k_1, \cdots$.
The duality simply amounts to permuting the order of these subspaces and
rearranging them into a new flag.

Let us consider a specific example of this duality in action.
The flag manifold $F(1,2,n)$ is the projectivization of the total space
of the tangent bundle to ${\bf P}^{n-1}$, and $F(1,n-1,n)$ is the
projectivization of the cotangent bundle to ${\bf P}^{n-1}$.
The projection to ${\bf P}^{n-1}$ in both cases is given by the forgetful
map to $F(1,n) = {\bf P}^{n-1}$, and the fibers of that projection
are $F(1,n-1)$ and $F(n-2,n-1)$, respectively, which are the projectivizations
of the tangent and cotangent spaces.  These two spaces are diffeomorphic:
one can be written in the form
\begin{displaymath}
\frac{U(n)}{U(1) \times U(1) \times U(n-2)}
\end{displaymath}
and the other in the form
\begin{displaymath}
\frac{U(n)}{U(1)\times U(n-2) \times U(1)}
\end{displaymath}
and there is certainly a diffeomorphism exchanging these two quotient spaces.
We can perform a consistency check of
the existence of that diffeomorphism by calculating cohomology
rings.
For a vector bundle $E \rightarrow M$ and rank $n$, the cohomology ring
of the total space of the projectivization of $E$ is given by
\cite{botttu}[section 20]:
\begin{displaymath}
H^*( {\bf P} E ) \: = \: H^*(M)[x]/(x^n + c_1(E)x^{n-1} +
\cdots + c_n(E) )
\end{displaymath}
The cohomology of the total spaces of the projectivization
of the tangent and cotangent bundles are isomorphic, related
in the notation above by sending
$x \mapsto -x$.
There exists an analogous diffeomorphism relating Hirzebruch surfaces
of different degree, though there the diffeomorphism preserves 
a projectivized complex vector bundle structure,
and so the diffeomorphism can be understood
by describing each Hirzebruch surface as the projectivization of a vector
bundle and arguing that the vector bundles are smoothly isomorphic,
modulo tensoring with an overall line bundle.
In the present case, by contrast, the diffeomorphism does not preserve
the projectivized complex vector bundle structure, 
and so no such argument can apply.
(In fact, such an argument necessarily cannot work.  One would need
a line bundle $L$ such that $T {\bf P}^{n-1} \otimes L \cong
T^* {\bf P}^{n-1}$ as smooth bundles, but this implies
\begin{displaymath}
c_1(T {\bf P}^{n-1}) \: + \: (n-1) c_1(L) \: = \: - c_1(T {\bf P}^{n-1} )
\end{displaymath}
or, more simply,
\begin{displaymath}
(n-1) c_1(L) \: = \: -2 c_1(T {\bf P}^{n-1}) \: = \: -2n
\end{displaymath}
but no such $L$ can exist, except possibly when $n-1=2$ or $1$.)

By repeatedly applying the duality above, one can show that
as a special case,
\begin{displaymath}
F(k_1, \cdots, k_n, N) \: \cong \: 
F(N-k_n, N-k_{n-1}, \cdots, N-k_2, N-k_1, N)
\end{displaymath}
Unlike the dualities above, this special case defines a
biholomorphism, not just a diffeomorphism.

It is straightforward to determine how Pl\"ucker coordinates behave
under the biholomorphic duality
\begin{displaymath}
F(k_1, \cdots, k_n, N) \: \leftrightarrow \: F(N-k_n, \cdots, N-k_1, N)
\end{displaymath}
of flag manifolds.
Let
\begin{displaymath}
\begin{array}{c}
p(AB\cdots C)_{i_1 \cdots i_{k_1}} \\
p(B\cdots C)_{i_1 \cdots i_{k_2}} \\
p(C)_{i_1 \cdots i_{k_n}}
\end{array}
\end{displaymath}
denote Pl\"ucker coordinates formed from the baryons $AB\cdots C$,
$B\cdots C$, and $C$ on the flag manifold $F(k_1, \cdots, k_n, N)$
above.
On the dual flag manifold,
let
\begin{displaymath}
\begin{array}{c}
q(AB\cdots C)_{i_1 \cdots i_{N-k_n}} \\
q(B\cdots C)_{i_1 \cdots i_{N-k_{n-1}}} \\
q(C)_{i_1 \cdots i_{N-k_1}}
\end{array}
\end{displaymath}
be their analogues.
Then, essentially\footnote{Contractions such as $AB \cdots C$
implicitly define projections to Grassmannians, so the duality described
here for Pl\"ucker coordinates on flag manifolds is, in fact,
an immediate consequence of the relation between Pl\"ucker coordinates
for dual Grassmannians.} as a consequence of the relation between 
Pl\"ucker coordinates on dual Grassmannians, the current sets of
Pl\"ucker coordinates are related in the form
\begin{displaymath}
p(AB\cdots C)_{i_1 \cdots i_{k_1}} \: = \: \frac{1}{(N-k_1)!}
\epsilon_{i_1 \cdots i_N} \, q(C)_{i_{k_1+1} \cdots i_N}
\end{displaymath}

In terms of the corresponding two-dimensional gauge theory,
the general duality means that a given $U(k_i)$ factor can be replaced by
a $U(k_{i+1}-k_i+k_{i-1})$ factor, at the same time that bifundamentals
are also replaced, and because the flag manifolds are the same,
for consistency of linear sigma model presentations one assumes that
the chiral rings of these two two-dimensional gauge theories also match.

This two-dimensional duality is very reminiscent of duality cascades
in four-dimensional gauge theories \cite{klebstrass,strasscas}.   
In particular, in the two-dimensional gauge theory,
the $U(k_i)$ factor has $k_{i+1} + k_{i-1}$ fundamentals,
and so applying Seiberg duality on that factor would replace
the $U(k_i)$ by $U(k_{i+1}-k_i + k_{i-1})$.  
Here, however, the analogy with Seiberg duality begins to break
down, because dualities on each individual $i$ do not typically
yield biholomorphisms, merely diffeomorphisms of the target space,
and unlike biholomorphisms, diffeomorphisms need not preserve the
chiral ring.

\subsection{Weighted Grassmannians}  \label{wtgrass}

There exists a notion of `weighted Grassmannians' \cite{cortireid},
which generalizes
both ordinary Grassmannians and weighted projective spaces.
Let us briefly review their construction and physical realization.

Briefly, the idea is to either modify the group $GL(k)$ ($U(k)$) appearing
in the GIT (symplectic) quotient construction of the Grassmannian
$G(k,N)$ with another group, and/or modify its action on the $N$ fundamentals.

Before describing the weighted Grassmannian, let us describe the
affine Grassmannian, as a simple prototype for these constructions.
The affine Grassmannian corresponding to the ordinary Grassmannian
$G(k,N)$ is defined by the GIT quotient
\begin{displaymath}
{\bf C}^{kN} // SL(k)
\end{displaymath}
 -- in other words, we replace $GL(k)$ by $SL(k)$.
The $SL(k)$ acts by sending a $k \times N$ matrix $A$ to $SA$,
for $S \in SL(k)$.
Physically, this is realized by an $SU(k)$ gauge theory with
$N$ fundamentals, instead of a $U(k)$ gauge theory with $N$ fundamentals.

A weighted Grassmannian can be constructed as a ${\bf C}^{\times}$
quotient of the affine Grassmannian.  However, that ${\bf C}^{\times}$
action does not lift to an action on ${\bf C}^{kN}$ -- 
a weighted Grassmannian is not, 
${\bf C}^{kN} // ( SL(k) \times {\bf C}^{\times})$.
Instead, to lift to an action on ${\bf C}^{kN}$, to describe the
weighted Grassmannian as a quotient of ${\bf C}^{kN}$ and so as a
nonabelian gauge theory, we must work a bit harder.
Specifically, we must replace ${\bf C}^{\times}$ by a cover
of ${\bf C}^{\times}$, call it $\widetilde{ {\bf C}^{\times} }$,
which {\it will} lift to an action on
${\bf C}^{kN}$, and then we must quotient out a noneffectively-acting
part of the $SL(k) \times \widetilde{ {\bf C}^{\times} }$ quotient.

The next step is to construct the group by which we will be quotienting
${\bf C}^{kN}$ to recover the weighted Grassmannian, not just the
ordinary Grassmannian.  In fact, we will see
there are several closely related groups,
all slightly different quotients of 
$SL(k) \times \widetilde{ {\bf C}^{\times}}$.

First, let us define $\widetilde{ {\bf C}^{\times} }$.
This is an extension of ${\bf C}^{\times}$ by ${\bf Z}_k$:
\begin{displaymath}
1 \: \longrightarrow \: {\bf Z}_k \: \longrightarrow \:
\widetilde{ {\bf C}^{\times} } \: \longrightarrow \: 
{\bf C}^{\times} \: \longrightarrow \: 1
\end{displaymath}
As a group, $\widetilde{ {\bf C}^{\times} } \cong {\bf C}^{\times}$,
but occasionally we need to distinguish them in order to 
precisely describe when a ${\bf C}^{\times}$ has a well-defined
action and when we have to replace it by a cover.
To be precise, and to hopefully help reduce confusion,
we will distinguish the two groups.

Next, let us define the gauge group.  This will be denoted $G_u$,
and is defined by
\begin{displaymath}
G_u \: = \: \frac{ SL(k) \times \widetilde{ {\bf C}^{\times} } }{ {\bf Z}_k }
\end{displaymath}
where the quotient is by the subgroup
\begin{displaymath}
\left\{ ( \zeta^u I, \zeta^{-1} ) \in SL(k) \times \widetilde{ {\bf C}^{\times}}
\, | \, \zeta^k = 1 \, \right\}
\end{displaymath}
for an integer $u$.  For different integers, we get different groups.
For example, when $u=0$, $G_0 = SL(k) \times {\bf C}^{\times}$.
For example, when $u=1$, $G_1 = GL(k)$.
These groups all have the same Lie algebra, but differ globally by
finite group factors.

Next, we need to describe how the group $G_u$ acts on
${\bf C}^{kN}$.  To do so, we will describe how
the cover $SL(k) \times \widetilde{ {\bf C}^{\times} }$ acts
on ${\bf C}^{kN}$.  That action will have a ${\bf Z}_k$ kernel,
and so will descend to an action of $G_u$.
So, let $A$ be a $k \times N$ matrix.
Let $S \in SL(k)$, and define a diagonal $N \times N$
matrix $D$ to have entries $\mu^{u + k w_i}$ on the diagonal,
for $\mu \in \widetilde{ {\bf C}^{\times} }$. 
Then, $SL(k) \times \widetilde{ {\bf C}^{\times} }$ acts as follows:
\begin{displaymath}
A \: \mapsto \: S A D
\end{displaymath}
This action has a ${\bf Z}_k$ center -- the action of the center of
$SL(k)$ is equivalent to multiplying on the right by a diagonal matrix.
In particular, for all $u$, this action descends to an action of $G_u$.

The weighted Grassmannian with weights $(u,w_1, \cdots, w_N)$ is
then the GIT quotient 
\begin{displaymath}
{\bf C}^{kN} // G_u
\end{displaymath}
with the 
$G_u$ action above.  In the case $u=1$ and all $w_i=0$,
the weighted Grassmannian reduces to the ordinary Grassmannian
$G(k,N)$.  In the special case $k=1$, the weighted Grassmannian
is the same as the weighted projective stack
${\bf P}^{N-1}_{[u+w_1, \cdots, u+w_N]}$.

Physically, this quotient is realized by a $\widetilde{G_u}$ gauge
theory, where $\widetilde{G_u}$ is a quotient of $SU(k) \times U(1)$
by a ${\bf Z}_k$ center of the same form as for $G_u$.
This gauge theory has $N$ chiral superfields transforming in  
$k$-dimensional representations of $G_u$, defined by the group action
defined above on ${\bf C}^{kN}$.

One can define baryons / Pl\"ucker coordinates in the same fashion described
earlier, as $SL(k)$-invariant field combinations.  Instead of defining an
embedding into an ordinary projective space, Pl\"ucker coordinates on 
a weighted Grassmannian now define an embedding into a weighted projective
space or stack, whose weights are given by 
$u + \sum(w_{\sigma(i)})$, summing over
all combinations of $k$ elements of the $N$ basis elements of ${\bf C}^N$.

We can also now shed some light on some earlier comments.
We claimed previously that a weighted Grassmannian is a 
${\bf C}^{\times}$ quotient of an affine Grassmannian.
That ${\bf C}^{\times}$ acts in such a way that the Pl\"ucker
coordinates have weights $u + \sum w_{\sigma(i)}$.
To lift that action to an action on ${\bf C}^{kN}$,
each of the $N$ sets of chiral superfields transforming as the
${\bf k}$ of $SL(k)$ would have to have weight $(u/k) + w_i$ under
the ${\bf C}^{\times}$.  Unless $u=0$ or is a multiple of $k$,
that does not make sense.  Instead, we replace ${\bf C}^{\times}$
with a $k$-fold cover, whose weights are $u + kw_i$, and then one
gets sensible results.

One can also define weighted flag manifolds in an analogous fashion,
though we shall not do so here.

\subsection{Mixed examples}

In this note we have studied gauged linear sigma models describing
flag manifolds.  It is also possible to mix flag manifolds and
toric varieties and stacks.  For completeness, we shall very briefly
outline a few examples here.

\subsubsection{ ${\bf P}^1$ bundle over flag manifold}

Let us describe a ${\bf P}^1$ bundle over a flag manifold
\begin{displaymath}
F(k_1, \cdots, k_n, N)
\end{displaymath}
Begin with a gauged linear sigma model for the flag manifold above,
{\it i.e.} a $U(k_1) \times \cdots \times U(k_n)$ gauge theory
with bifundamental matter,
and add two more chiral superfields $p_0$, $p_1$.
Let $p_1$ be neutral under the determinants of each $U(k_i)$,
but let $p_0$ be charged under those same determinants.
Then, if we gauge an additional $U(1)$ that rotates $p_0$, $p_1$
by the same phase factors, then we have a ${\bf P}^1$ bundle
over the flag manifold, built as a projectivization ${\bf P}{\cal E}$ of 
a rank two vector bundle ${\cal E} = {\cal O} + L$ on the flag manifold.

Similarly, one can fiber other toric varieties and stacks over a flag manifold.

\subsubsection{Gerbe on a flag manifold}

Let us describe a ${\bf Z}_k$ gerbe over the flag manifold
$F(k_1, \cdots, k_n, N)$ above.
Begin as above with a gauged linear sigma model for the flag manifold,
then add one extra chiral superfield $p$ with charge $q_i$ under
$\mbox{det }U(k_i)$.
Now, gauge an additional $U(1)$ that acts solely on $p$, with charge $m$.

The D-term for the additional $U(1)$ has the form
\begin{displaymath}
k |p|^2 \: = \: r
\end{displaymath}
and so when $r \gg 0$, we see that $p \neq 0$, and so $p$ is describing
a ${\bf C}^{\times}$ bundle over the flag manifold.
Gauging $U(1)$ rotations (in the supersymmetric theory) removes all
physical degrees of freedom along the fiber directions.
Giving $p$ charge $m$ rather than charge $1$ means that we are
`overgauging', gauging $m$ rotations rather than a single rotation.
This distinction is nonperturbatively meaningful, as discussed
in \cite{stxglsm}, and the resulting low-energy theory is a sigma
model on a ${\bf Z}_m$ gerbe over the flag manifold, with characteristic
class
\begin{displaymath}
( q_1 \mbox{ mod }m, q_2 \mbox{ mod }m, \cdots, q_n \mbox{ mod }m)
\end{displaymath}

\subsubsection{Grassmannian bundle over ${\bf P}^1$}

The simplest possible example of a fibered flag manifold
is a ${\bf P}^1$ bundle over ${\bf P}^1$, {\it i.e.}
Hirzebruch surfaces.
In terms of gauged linear sigma models, if we let 
$u$, $v$ be homogeneous coordinates on the base and
$s$, $t$ be homogeneous coordinates on the fibers,
then we can build such Hirzebruch surfaces as $U(1)^2$ gauge theories
with charges
\begin{center}
\begin{tabular}{cccc} 
$u$ & $v$ & $s$ & $t$ \\ \hline
$1$ & $1$ & $n$ & $0$ \\
$0$ & $0$ & $1$ & $1$ 
\end{tabular}
\end{center}

Such notions can be easily generalized.  For example, we can describe
a Grassmannian $G(k,N)$ bundle over ${\bf P}^1$, as follows.
This will be a $U(1) \times U(k)$ gauge theory, with matter
fields $u$, $v$ (forming homogeneous coordinates on the base ${\bf P}^1$)
and $\phi^{is}$ ($i \in \{ 1, \cdots, k\}$, $s \in \{ 1, \cdots, N \}$,
forming the Grassmannian fiber).
The $U(k)$ only acts on the $\phi^{is}$, which transform in $N$ copies
of the fundamental representation.
The $U(1)$ acts on $u$, $v$ with charge 1, and also acts on the 
$\phi^{is}$ with weight $p_s$, where $p_s$ is a sequence of $N$ integers.
(The local $U(k)$ gauge symmetry is preserved but the global $U(N)$
symmetry is broken.)
More invariantly, we can think of this as a bundle with fibers
$G(k, {\cal E})$ where
\begin{displaymath}
{\cal E} \: = \: \oplus_{s=1}^N {\cal O}_{{\bf P}^1}(p_s)
\end{displaymath}
More general examples are also certainly possible, though this should
suffice for illustrative purposes.

\section{Bundles}   \label{flagbdles}

\subsection{Tangent bundles and the (2,2) locus}

First, let us recall how the tangent bundle is described by left-moving
fermions in the linear sigma model for ${\bf P}^{N-1}$.
Recall that ${\bf P}^{N-1}$ is described by a theory of $N$ chiral
fields in which a $U(1)$ action has been gauged.
The action for the gauged linear sigma model has interaction terms of the
form  
\begin{displaymath}
\overline{\phi}_i \psi_-^i \lambda_+ \: - \:
\overline{\phi}_i \psi_+^i \lambda_- \: + \: {\it c.c.}
\end{displaymath}
where $\phi$ are the bosonic parts of the $N$ chiral multiplets,
$\psi$ their superpartners, and $\lambda$ part of the gauge multiplet.
When $\phi$ has a nonzero vev, we see a linear combination of $\psi$'s
and $\lambda$'s becomes massive.  To be specific, consider the term
containing $\psi_+$, and consider the subset of $\psi_+^i$'s defined by
$\psi_+^i \: = \: \psi \phi_i$ for some Grassmann-valued parameter $\psi$.
Using the D-term condition
\begin{displaymath}
\sum_i | \phi_i |^2 \: = \: r
\end{displaymath}
we see that the Yukawa coupling involving this particular $\psi_+^i$
reduces to
\begin{displaymath}
\overline{\phi}_i \psi_+^i \lambda_- \: = \: 
\psi \lambda_- | \phi_i |^2 \: = \: r \psi \lambda_-
\end{displaymath}
Thus, it is the $\psi \phi_i$ combination that becomes massive.
Put another way, identifying the $\psi_+^i$'s with local sections of
${\cal O}(1)^{\oplus N}$, we see that the remaining massless $\psi_+^i$'s
couple to the cokernel $T$ in the short exact sequence below:
\begin{displaymath}
0 \: \longrightarrow \: {\cal O} \: \stackrel{ \otimes \phi_i }{\longrightarrow}
\: {\cal O}(1)^{\oplus N} \: \longrightarrow \: T \: \longrightarrow \:
0
\end{displaymath}
but that cokernel $T$ is the tangent bundle of ${\bf P}^{N-1}$,
so we see that the remaining massless $\psi^i$'s couple to the tangent
bundle.

Mathematically, on ${\bf P}^{N-1}$ we have the ``universal subbundle''
(or ``tautological bundle'')
$S$, of rank one and $c_1(S) = -1$, specifically
${\cal O}(-1)$, and the ``universal quotient bundle''
$Q$ of rank $N-1$ and $c_1(Q) = +1$.  These are related by the short
exact sequence
\begin{displaymath}
0 \: \longrightarrow \: S \: \longrightarrow \: {\cal O}^N \:
\longrightarrow \: Q \: \longrightarrow \: 0
\end{displaymath}
The tangent bundle of the projective space is given by
$S^{\vee} \otimes Q$, and so fits into the short exact sequence
\begin{displaymath}
0 \: \longrightarrow \: {\cal O} \: \longrightarrow \:
{\cal O}(1)^{\oplus N} \: \longrightarrow \: T {\bf P}^{N-1}
\: \longrightarrow \: 0
\end{displaymath}
known as the ``Euler sequence.''

Projective spaces are special examples of Grassmannians -- specifically,
\begin{displaymath}
{\bf P}^{N-1} \: = \: G(1,N)
\end{displaymath}
in our notation.  Thus, it should not be surprising that the analysis 
above
generalizes easily to Grassmannians.

Let us begin our description of Grassmannians by working through
the relevant mathematics.
On any Grassmannian $G(k,N)$, we have the ``universal subbundle''
$S$, of rank $k$ and $c_1(S)=-1$,
defined at any point on the Grassmannian to have fiber defined by
the $k$-dimensional subspace of ${\bf C}^N$ corresponding to that point.
In addition, we have
the ``universal quotient bundle'' $Q$ of rank $N-k$ and
$c_1(Q)=+1$.  Let $V$ denote the trivial rank $N$ bundle on
$G(k,N)$, then these are related by the short exact sequence
\begin{displaymath}
0 \: \longrightarrow \: S \: \longrightarrow \: V \: \longrightarrow
\: Q \: \longrightarrow \: 0
\end{displaymath}
Under the duality operation $G(k,N) \leftrightarrow G(N-k,N)$,
the universal subbundle $S$ is exchanged with $Q^{\vee}$,
the dual of the universal
quotient bundle, and {\it vice-versa}.
The tangent bundle $T = \mbox{Hom}(S,Q) = S^{\vee} \otimes Q$,
and so can be described as the cokernel
\begin{displaymath}
0 \: \longrightarrow \: S^{\vee} \otimes S \: \longrightarrow \:
S^{\vee} \otimes V \: \longrightarrow \: T \: \longrightarrow \: 0
\end{displaymath}
(For later use, note that using multiplicative properties of Chern
characters ($\mbox{ch}(T) = \mbox{ch}(S^{\vee})\mbox{ch}(Q)$)
one can immediately show $c_1(T) = N$.)

The analysis of the physics for Grassmannians also proceeds much as for 
projective
spaces.  The lagrangian of the corresponding nonabelian
GLSM contains Yukawa couplings of the form
\begin{displaymath}
\overline{\phi}_{is} \lambda_{- \, j}^i \psi_+^{js}
\end{displaymath}
(plus complex conjugates and other chiralities)
where $i$ is a $U(k)$ index and $s \in \{ 1, \cdots, N \}$.
Thus, some combination of $\lambda_-$'s and $\psi_+$'s will become massive.
The $\psi_+$'s that become massive can be understood by making the
ansatz $\psi_+^{is} = \psi \phi_{is}$, then applying the D-terms we see
\begin{displaymath}
\overline{\phi}_{is} \lambda_{- \, j}^i \psi_+^{js}
\: = \:
\overline{\phi}_{is} \lambda_{- \, j}^i \phi^{js} \psi
\: = \: r \delta^j_i \lambda_{- \, j}^i \psi \: = \:
r \lambda_{- \, i}^i \psi
\end{displaymath}
and so we see that it is the $\psi \phi_{is}$ combination of $\psi_+$'s
that becomes massive.  Put another way, the remaining massless
$\psi_+$'s couple to the cokernel $T$
of a short exact sequence of the form
\begin{displaymath}
0 \: \longrightarrow \: S^{\vee}\otimes S \: \stackrel{ \otimes \phi_{is} }{
\longrightarrow} \: S^{\vee} \otimes V \: \longrightarrow \: T
\: \longrightarrow \: 0
\end{displaymath}
and so we realize the tangent bundle $T$ of the Grassmannian in physics.

The tangent bundle of a partial flag manifold can be realized similarly.
As for Grassmannians, let us first describe the tangent bundle mathematically,
then we shall describe how it is realized physically in the corresponding
nonabelian GLSM.

Mathematically, the main difference between the tangent bundle of
a Grassmannian and that of a more general partial flag manifold
is that instead of a single universal subbundle and universal
quotient bundle, we now have a flag of both.
First, over $F(k_1, \cdots, k_n, N)$ there is a universal flag of
subbundles
\begin{displaymath}
S_1 \: \hookrightarrow \: S_2 \: \hookrightarrow \: \cdots \:
\hookrightarrow \: S_n \: \hookrightarrow \: V
\end{displaymath}
where the rank of $S_i$ is $k_i$, and $V$ is the trivial rank $N$ bundle
over the flag manifold.
(The fibers of these subbundles are defined at any point on the flag
manifold to be the flag corresponding to that point.)
In addition, there is a dual flag of quotient bundles
\begin{displaymath}
V \: \longrightarrow \: Q_1 \: \longrightarrow \: Q_2 \: \longrightarrow
\: \cdots \: \longrightarrow \: Q_{n} 
\end{displaymath}
where each of the maps above is onto, and the $Q_i$'s are defined by
the short exact sequences
\begin{displaymath}
0 \: \longrightarrow \: S_i \: \longrightarrow \: V \: \longrightarrow
\: Q_i \: \longrightarrow \: 0
\end{displaymath}
Furthermore, the classical cohomology ring of the flag manifold
is generated by the Chern classes of the quotients $S_i/S_{i-1}$.

The tangent bundle is constructed as a sequence of extensions of 
bundles $T_i$ defined as the following cokernels:
\begin{displaymath}
0 \: \longrightarrow \: S_i^{\vee} \otimes S_i \: \longrightarrow
S_i^{\vee} \otimes S_{i+1} \: \longrightarrow \: T_i \: 
\longrightarrow \: 0
\end{displaymath}
in conventions where $S_{n+1} = V$.
Note that as a special case, when $n=1$, so that the flag variety reduces
to a Grassmannian, the tangent bundle $T_1$ is the same as that of
the Grassmannian.
Now, as a smooth bundle, the tangent bundle of the flag manifold
is isomorphic to the
direct sum over the $T_i$'s, though the holomorphic structure is a bit
more complicated.  Holomorphically, 
this is an extension rather than a direct sum.
Consider the partial flag varieties $F_i = F(k_i, \cdots, k_n, N)$,
and let $P_i$ be the pullback to $F$ of the tangent bundle to $F_i$.
This gives a series of extensions relating the intermediate bundles
$P_i$:
\begin{displaymath}
\begin{array}{c}
P_n \: = \: T_n \\
0 \: \longrightarrow \: T_{n-1} \: \longrightarrow \: P_{n-1} \:
\longrightarrow \: P_n \: \longrightarrow \: 0 \\
0 \: \longrightarrow \: T_{n-2} \: \longrightarrow \: P_{n-2}
\: \longrightarrow \: P_{n-1} \: \longrightarrow \: 0 \\
\vdots \\
0 \: \longrightarrow \: T_1 \: \longrightarrow \: P_1 \:
\longrightarrow \: P_2 \: \longrightarrow \: 0 \\
P_1 \: = \: TF
\end{array}
\end{displaymath}
where $TF$ is the tangent bundle of the flag manifold.
More formally, the $T_i$ form the associated graded bundles to the
tangent bundle, and the extensions above reconstruct the tangent bundle
from its associated graded.

The tangent bundle $TF$ of the flag manifold can also be described more
compactly as the cokernel of the short exact sequence
\begin{displaymath}
0 \: \longrightarrow \: 
\bigoplus_{i=1}^n S_i^{\vee} \otimes S_i 
\: \stackrel{ * }{\longrightarrow } \:
\bigoplus_{i=1}^n S_i^{\vee} \otimes S_{i+1} \:
\longrightarrow \: TF \: \longrightarrow \: 0
\end{displaymath}
where the map $*$ consists of the bifundamentals $\phi: S_i \rightarrow S_{i+1}$
on the diagonal and $\phi^{\vee}: S_{i+1}^{\vee} \rightarrow S_i^{\vee}$
just above the diagonal.  For example, for the flag manifold $F(k_1,k_2,N)$,
the map $*$ is given by
\begin{displaymath}
\left[ \begin{array}{cc} 
\phi_{12} & \phi_{12}^{\vee} \\
0 & \phi_{23} 
\end{array} 
\right]
\end{displaymath}
where $\phi_{ij}: S_i^{\vee} \otimes S_i \rightarrow
S_i^{\vee} \otimes S_j$ is the map induced by the bifundamental
mapping $S_i \rightarrow S_j$, and 
$\phi^{\vee}_{ij}: S_j^{\vee}\otimes S_j \rightarrow
S_i^{\vee} \otimes S_j$ is the map induced by the dual bifundamental
$S_j^{\vee} \rightarrow S_i^{\vee}$.
In other words,
\begin{eqnarray*}
(\phi_{12}):  S_1^{\vee} \otimes S_1  & \longrightarrow &
S_1^{\vee} \otimes S_2 \\
(\phi_{12}^{\vee} + \phi_{23}):  S_2^{\vee} \otimes S_2  & \longrightarrow &
( S_1^{\vee} \otimes S_2 ) \oplus ( S_2^{\vee} \otimes S_3 )
\end{eqnarray*}
Similarly, for the flag manifold $F(k_1,k_2,k_3,N)$, the map $*$ is given
by
\begin{displaymath}
\left[ \begin{array}{ccc}
\phi_{12} & \phi_{12}^{\vee} & 0 \\
0 & \phi_{23} & \phi_{23}^{\vee} \\
0 & 0 & \phi_{34}
\end{array} \right]
\end{displaymath}
We see that the quotient $TF$ is filtered but, since the maps are not
block diagonal, it does not decompose as a direct sum.
It can be deformed to the direct sum of the $T_i$ by setting the
off-diagonal terms to 0.

Next, we shall describe how this structure is realized physically in
a nonabelian GLSM describing the partial flag manifold.
Consider the flag manifold $F(k_1, \cdots, k_n, N)$, realized
as a two-dimensional $U(k_1) \times \cdots 
\times U(k_n)$ gauge theory with bifundamentals.
First consider the $U(k_1)$ factor.
There are Yukawa couplings of the form
\begin{displaymath}
\overline{\phi}_{is} \lambda_{- \, j}^i \psi_+^{js}
\end{displaymath}
(where $s$ is a $U(k_2)$ index), which can be analyzed in
exactly the same way as those for a Grassmannian.
The combination $\psi_+^{is} = \psi \phi^{is}$ becomes massive.
Next, suppose $\psi_+^{is}$ are the fermionic components of a chiral
superfield in the $({\bf k_{i-1}}, {\bf \overline{k_{i}}})$
representation of $U(k_{i-1}) \times U(k_{i})$,
and $\tilde{\psi}_+^{ai}$ are the corresponding components of a chiral
superfield in the $({\bf k_i}, {\bf \overline{k_{i+1}}})$
representation of $U(k_i) \times U(k_{i+1})$.
Here, the Yukawa couplings involving the $U(k_i)$ $\lambda_-$'s are of the form
\begin{displaymath}
- \overline{\phi}_{is} \lambda_{- \, j}^i \psi_+^{js} \: + \:
\overline{\tilde{\phi}}_{ai} \lambda_{- \, j}^i \tilde{\psi}^{a j}
\end{displaymath}
If we make the ansatz $\psi_+^{is} = \psi \phi^{is}$ and
$\tilde{\psi}_+^{ai} = \psi \tilde{\phi}^{ai}$, 
for the same Grassmann parameter $\psi$ in both cases,
then using the
D-terms the Yukawa couplings 
above reduce to 
\begin{displaymath}
\left( - \overline{\phi}_{is} \phi^{js} \: + \:
\overline{\tilde{\phi}}_{ai} \tilde{\phi}^{aj} \right) \lambda_{- \, j}^i
\psi \: = \: r_i \mbox{Tr }\lambda_- \psi
\end{displaymath}
Putting this together, we see that 
the remaining massless $\tilde{\psi}_+$'s are described
by the cokernel of the short exact sequence
\begin{displaymath}
0 \: \longrightarrow \: 
\bigoplus_{i=1}^n S_i^{\vee} \otimes S_i 
\: \stackrel{ * }{\longrightarrow } \:
\bigoplus_{i=1}^n S_i^{\vee} \otimes S_{i+1} \:
\longrightarrow \: TF \: \longrightarrow \: 0
\end{displaymath}
where the $\oplus S_i^{\vee} \otimes S_i$ factor is realized by the
gauginos $\lambda$, the $\oplus S_i^{\vee} \otimes S_{i+1}$ factor is
realized by the fermionic parts $\psi$ of the bifundamentals,
and $*$ is the matrix with bifundamentals on and just above the diagonal
described earlier.
Thus, we see that the physical analysis of the fermions matches
the mathematical picture of $TF$ described earlier.

\subsection{Aside:  homogeneous bundles}

For any parabolic subgroup $P$ of a reductive algebraic group $G$,
we can construct a large number of bundles on the coset space
$G/P$ by using a representation $\rho$ of $P$.
The total space of the bundle is
$G \times_{\rho} V$, which has a natural projection to $G/P$, 
where $V$ is the vector space
on which the representation $\rho$ acts.  Such bundles are known as
``homogeneous bundles.''

Grassmannians and flag manifolds can be described as coset spaces $G/P$,
where in each case $G = GL(N)$, and their tangent bundles are examples
of homogeneous bundles.
The representations of the relevant parabolic $P$ for a Grassmannian
$G(k,N)$ are the
same as those of $U(k) \times U(N-k)$, so we merely need to find
the relevant representation of $U(k) \times U(N-k)$.
The universal subbundle $S$ corresponds to the fundamental
representation ${\bf k}$ of $U(k)$, and the universal quotient bundle
$Q$ corresponds to the dual ${\bf \overline{N-k}}$ of the fundamental
representation of $U(N-k)$.  Since the tangent bundle of $G(k,N)$
is given by $\mbox{Hom}(S,Q)$, we see that the tangent bundle is 
defined in this fashion by the representation
$({\bf \overline{k}}, {\bf \overline{N-k}})$ of $U(k) \times U(N-k)$.

The tangent bundle of the flag manifold can be described in an
analogous fashion.  
Each of the bundles $S_i$ can be described by the representation
\begin{displaymath}
\left( {\bf k_1}, 0, \cdots, 0 \right) \oplus
\left( 0, {\bf k_2 - k_1}, 0 \cdots, 0 \right)
\oplus \left( 0, 0, {\bf k_3 - k_2}, 0, \cdots, 0 \right)
\oplus \left( 0, \cdots, 0, 
{\bf k_i - k_{i-1}}, 0, \cdots, 0 \right)
\end{displaymath}
of $U(k_1) \times U(k_2 - k_1) \times \cdots \times U(N - k_n)$.
Call this representation $\rho_{S_i}$.
Similarly,
each of the bundles $T_i$ that appeared in the
construction of the tangent bundle is homogeneous, and from their
definition it should be clear that they are defined by
the representation 
\begin{eqnarray*}
\lefteqn{\rho_{S_i}^* \otimes \left( 0, \cdots, 0, {\bf \overline{ k_{i+1} - k_i
 }},
0, \cdots, 0 \right) } \\
& & \: = \: \left( {\bf \overline{k_1}}, 0, \cdots, 0,
{\bf \overline{k_{i+1}-k_i}}, 0, \cdots, 0 \right) \oplus \cdots
\oplus \left( 0, \cdots, 0, {\bf \overline{k_i - k_{i-1}}},
{\bf \overline{ k_{i+1}-k_i}}, 0, \cdots, 0\right)
\end{eqnarray*}

However, typical bundles one works with in a (0,2) model will not
be homogeneous, as we shall see in the next section.

\subsection{More general bundles}

It is straightforward to describe other bundles in (0,2) GLSM's over
flag manifolds.  For example, one common type of bundle described
with (0,2) GLSM's is given as the kernel ${\cal E}$ of a short
exact sequence:
\begin{displaymath}
0 \: \longrightarrow \: {\cal E} \: \longrightarrow \:
{\cal V}_1 \: \stackrel{F}{\longrightarrow} \: {\cal V}_2 \:
\longrightarrow \: 0
\end{displaymath}
The bundles ${\cal V}_1$ and ${\cal V}_2$ are determined by
representations $\rho_1$, $\rho_2$ of the gauge group.
Physically, to describe ${\cal V}_1$ we introduce left-moving
fermi multiplets $\Lambda$ in the $\rho_1$ representation of the
gauge group, and right-moving chiral superfields $p$ in the
$\rho_2^*$ representation of the gauge group.  The map
${\cal V}_1 \rightarrow {\cal V}_2$ is realized by a
(0,2) superpotential
\begin{displaymath}
W \: = \: \int d \theta p F \Lambda
\end{displaymath}
For example, one of the terms descending from that superpotential is
\begin{displaymath}
\psi_p^s \lambda_-^i F^{is}(\phi)
\end{displaymath}
So long as the $\phi$'s have nonzero vevs, a linear combination of
$\psi_p$ and $\lambda_-$ will become massive.
The subset of $\lambda_-$ that remain massless are given by the
kernel of the map defined by $F$, hence this superpotential describes
a kernel.

For example, consider a bundle on $G(k,N)$ built as a kernel.
Let $\rho_1$ be ${\bf k}$, and $\rho_2$ be invariant under $SU(k)$,
Take
\begin{displaymath}
F^{i} \: = \: \sum_{s=1}^N \alpha_s \phi^{is}
\end{displaymath}
for some arbitrary constants $\alpha_i$.
With the superpotential $\int d\theta p \Lambda F$, we have a (0,2)
GLSM on $G(k,N)$ with bundle described by the kernel of $F$ above.
Note that this bundle is not homogeneous, and cannot be described in terms
of representation theory.

For another example, let us build a bundle on $G(2,N)$ as a kernel.
Take $\rho_1$ to be the ${\bf 2} \otimes {\bf 2}$ representation
of $U(2)$, and take $\rho_2$ to be invariant under $SU(2)$.
Take the map $F^{ij} = \epsilon^{ij}$.  Then our bundle over
$G(2,N)$ is defined by the symmetric (${\bf 3}$)
representation of $U(2)$.

Other standard (0,2) bundle constructions \cite{distrev}
are also possible, but we shall
not discuss them further here.

\section{Complete intersections in flag manifolds}
\label{ciflag}

\subsection{Calabi-Yau's in flag manifolds}

We can add matter fields $p_{\alpha}$ to describe hypersurfaces with
gauge-invariant superpotentials, following \cite{WitPhases,hori1}.
The condition for the intersection of those hypersurfaces to
be Calabi-Yau can be described as the condition for the axial anomaly
in the two-dimensional $(2,2)$ gauged linear sigma model to vanish.
In effect, there will be $n$ $U(1)$ factors, corresponding to the
determinants of the $U(k_i)$'s, and so $n$ conditions for the
intersection to be Calabi-Yau.
Following the analysis of \cite{hori1} in the Grassmannian case,
and given that our gauged linear sigma models for flag manifolds
have bifundamental matter, it is straightforward to see that
the Calabi-Yau condition can be stated as follows:
\begin{center}
\begin{tabular}{cc}
$i$ & Minus the sum of $p_{\alpha}$ charges in det $U(k_i)$ ($=-K_F$)\\ \hline
$1$ & $k_2$ \\
$2$ & $k_3 - k_1$ \\
$3$ & $k_4 - k_2$ \\
$\cdots$ & $\cdots$ \\
$n$ & $N-k_{n-1}$
\end{tabular}
\end{center}
The right column is determined as the difference between the number of
fundamentals and antifundamentals charged under the corresponding
$U(k_i)$.  The right column can also be understood mathematically
as minus the part of the canonical divisor corresponding to that
$U(k_i)$ factor.

A list of Calabi-Yau three-folds obtained as complete intersections
in flag manifolds is provided at the end of \cite{batyrev2}.
For completeness, we repeat that list here:
\begin{center}
\begin{tabular}{cccc}
$n$ & $F$ & $\mbox{dim}(F)$ & $-K_F$ \\ \hline
$7$ & $F(2,7)$ & $10$ & $7$ \\
$7$ & $F(1,2,7)$ & $11$ & $(2,6)$ \\
$7$ & $F(1,5,7)$ & $14$ & $(5,6)$ \\
$7$ & $F(1,2,6,7)$ & $15$ & $(2,5,5)^*$ \\
$6$ & $F(2,6)$ & $8$ & $6$ \\
$6$ & $F(3,6)$ & $9$ & $6$ \\
$6$ & $F(1,2,6)$ & $9$ & $(2,5)$ \\
$6$ & $F(1,3,6)$ & $11$ & $(3,5)$ \\
$6$ & $F(1,4,6)$ & $11$ & $(4,5)$ \\
$6$ & $F(1,2,5,6)$ & $12$ & $(2,4,4)$ \\
$6$ & $F(1,3,5,6)$ & $13$ & $(3,4,3)$ \\
$5$ & $F(2,5)$ & $6$ & $5$ \\
$5$ & $F(1,2,5)$ & $7$ & $(2,4)$ \\
$5$ & $F(2,3,5)$ & $8$ & $(3,3)$ \\
$5$ & $F(1,3,5)$ & $8$ & $(3,4)$ \\
$5$ & $F(1,2,4,5)$ & $9$ & $(2,3,3)$ \\
$5$ & $F(1,2,3,5)$ & $9$ & $(2,2,3)$ \\
$5$ & $F(1,2,3,4,5)$ & $10$ & $(2,2,2,2)$ \\
$4$ & $F(2,4)$ & $4$ & $4$ \\
$4$ & $F(1,2,4)$ & $5$ & $(2,3)$ \\
$4$ & $F(1,2,3,4)$ & $6$ & $(2,2,2)$
\end{tabular}
\end{center}
(*:  This corrects a trivial typo in \cite{batyrev2}.)

Note that the Calabi-Yau's built in Grassmannians listed in
\cite{hori1}[table 2] and \cite{batyrev1}[section 6.2]
form a subset of the list above.

The observant reader will note that,
according to our earlier description of dualities,
\begin{displaymath}
\begin{array}{c}
F(1,3,6) \: \cong \: F(1,4,6) \\
F(1,3,5) \: \cong \: F(2,3,5) \\
F(1,2,4,5) \: \cong \: F(1,2,3,5)
\end{array}
\end{displaymath}
should be related by diffeomorphisms.  However, although the
manifolds are diffeomorphic, note that $-K_F$ differs.
This is consistent because $-K_F$ is determined in part by the
complex structure, and the diffeomorphisms relating the special
cases above do not preserve the complex structure.
Put another way, $c_1(K)$ matches $c_1$ of the holomorphic part of the
complexified tangent bundle, but as Chern classes can only be defined
for complexified tangent bundles, there is no reason why the
$c_1$'s need be invariant under non-holomorphic diffeomorphisms.
For a simpler example of this principle, the map $z \mapsto \overline{z}$
defines a non-holomorphic diffeomorphism ${\bf P}^1 \rightarrow {\bf P}^1$,
which sends $c_1(K) \mapsto - c_1(K)$.

The analysis also easily extends to the weighted Grassmannians and
flag manifolds described in section~\ref{wtgrass}.
For example, for a weighted Grassmannian modelled on $G(k,N)$
with weights $(u,w_1, \cdots, w_N)$, the Calabi-Yau condition for a complete
intersection is that the sum of the degrees of the hypersurfaces
must equal
\begin{displaymath}
N u \: + \: k \sum_i w_i
\end{displaymath}
For an ordinary Grassmannian $G(k,N)$, $u=1$ and all the
$w_i = 0$,
and so the condition reduces to the constraint that the sum of the
degrees equal $N$.

\subsection{Non-birational derived equivalences in nonabelian GLSMs}

In \cite{ps4}[section 12.2] and \cite{hori1}, examples were given of
GLSM's in which 
\begin{itemize}
\item one of the K\"ahler phases had a geometric interpretation, realized
in a novel fashion, and
\item geometric K\"ahler phases were not birational
\end{itemize}
in abelian and nonabelian GLSM's, respectively.
It has long been assumed that the geometric phases of GLSM's were
related by birational transformations, so examples contradicting that
belief are of interest.

In this section we shall elaborate on this matter, by studying further
examples of this phenomenon in nonabelian GLSM's, and reviewing
an example of the phenomenon in abelian GLSM's.
We will also describe a tentative proposal for understanding the
mathematical relationship between non-birational phases:
we propose that they should be understood in terms of Kuznetsov's
homological projective duality \cite{kuz1,kuz2,kuz3}.
In this paper we will only begin to outline the relevance of Kuznetsov's
work -- a much more thorough description, and further application to
abelian GLSM's, will appear in \cite{alltoappear}.

We shall begin by reviewing the example in \cite{hori1}, then discuss
other examples in nonabelian GLSM's before interpreting the results
in terms of Kuznetsov's work and outlining an analogous example in
abelian GLSM's.

\subsubsection{Hori-Tong-Rodland example}   \label{horitongrodland}

In \cite{hori1}[section 5], an example of a gauged linear sigma model
was analyzed which described, for $r \gg 0$, a complete intersection
of seven degree one hypersurfaces in the Grassmannian
$G(2,7)$, and for $r \ll 0$, the ``Pfaffian Calabi-Yau,'' which
is not a complete intersection.  (See \cite{andreilev} for 
a corresponding mathematical discussion.)
The superpotential terms each
involve two of the chiral fields $\Phi^a_i$ 
defining the Grassmannian and a single
auxiliary field $p^i$, introduced to describe members of the complete
intersection, and so has the form
\begin{displaymath}
W \: \propto \: A^{jk}_i p^i \Phi^a_j \Phi^b_k \epsilon_{ab}
\end{displaymath}
for constants $A^{jk}_i$.
In terms of the $U(1) \subset U(2)$ given by matrices proportional to the
identity, the $p^i$ have charge 2 and the $\Phi^a_i$ have charge 1.
For $r \gg 0$, it is straightforward to check that the gauged linear
sigma model describes $G(2,7)[1^7]$.
For $r \ll 0$, the D-terms forbid the $p^i$ from all vanishing,
and so they form homogeneous coordinates on a ${\bf P}^6$.
The $U(2)$ gauge symmetry is broken to $SU(2) \times {\bf Z}_2$,
where the ${\bf Z}_2$ subgroup of $U(2)$ is given by 
matrices of the form $\mbox{diag}(1,\zeta)$, for $\zeta = \pm 1$.
That ${\bf Z}_2$ acts trivially on the $p^i$ but acts effectively
on the $\Phi^a_i$.
The matrix $A^{jk}_i p^i \equiv A(p)^{jk}$ is a skew-symmetric $7 \times 7$
matrix with entries
linear in the $p^i$.  Nondegeneracy of the original complete intersection
forces $A(p)$ to have rank either $4$ or $6$ for all $p$.  
At $p$ for which $A(p)$ has rank $4$, there are $7-4=3$ massless $\Phi$
doublets; the rest are massive with masses proportional to $|r|$.
At $p$ for which $A(p)$ has rank $6$, there is $7-6=1$ massless
$\Phi$ doublet.
As discussed in \cite{hori1}, in the infrared limit all vacua in the
nonabelian gauge theory on the rank $6$ locus run to infinity,
whereas on the rank $4$ locus a single vacuum remains.
Thus, \cite{hori1} identify the $r \ll 0$ phase with a nonlinear sigma
model on the vanishing locus of the $6 \times 6$ Pfaffians of
the matrix $A(p)$ defined over ${\bf P}^6$.

\subsubsection{Aside:  Pfaffian varieties}   \label{revpfaff}

As Pfaffian varieties are not commonly used in the physics
community, let us take a moment to check certain basic properties
of the previous example.
(For a more complete description, see for example 
\cite{hubsch}[section 3.5].  For other related information,
see for example \cite{harristu}.)

First, let us establish some notation that will sometimes be used
elsewhere in this paper.
The Pfaffian variety $Pf(N)$ for $N$ odd is a space defined as follows.
Let $V$ be an $N$-dimensional vector space, and consider the locus
in the space of two-forms in $\Lambda^2 V^*$ whose rank is not maximal.
The maximal possible rank is $N-1$, and the next possible rank is $N-3$.
So, for $N$ odd, $Pf(N)$ can be defined as the space of skew-symmetric
$N \times N$ matrices of rank $N-3$, {\it i.e.} whose $(N-1)\times (N-1)$
Pfaffians all vanish.

In particular, $Pf(5)$ and $G(2,5)$ are the same space, but on
dual vector spaces.  In terms of Pl\"ucker coordinates,
$G(2,5)$ describes the space of elements of $\Lambda^2 V$ which have
the smallest possible nonzero rank, whereas $Pf(5)$ describes the
locus of elements of $\Lambda^2 V^*$ of submaximal rank, which is also
2 in this case.  Alternatively, by identifying matrix entries with coordinates
on the space of all skew-symmetric matrices and performing a linear change of
coordinates, we can describe $Pf(5)$ (and hence $G(2,5)$) as the vanishing
locus in ${\bf P}^{9}$ (the projectivized space of all skew-symmetric
$5 \times 5$ matrices, using the fact that such matrices have ten different
entries) of the $4 \times 4$ Pfaffians of a generic $5 \times 5$ skew-symmetric
matrix whose entries are linear in the homogeneous coordinates on ${\bf P}^9$.

For $N > 5$, the picture is slightly richer.  The group
$GL(V) = GL(N)$ acts on ${\bf P}(\Lambda^2 V^{\vee})$ with a finite
number of orbits ${\it O}_{2k}$, classified by an even
integer $2k \leq N$.  Orbit ${\it O}_{2k}$ consists of
(the projectivizations of) all skew-symmetric matrices $A$ of rank equal
to $2k$.  (Clearly, any such $A$ can be taken to any other by some element
of $GL(V)$.)  In general, the Grassmannian is
$G(2,V) = {\it O}_2$, while the Pfaffian is $Pf = {\it O}_{N-3}$ for
$N$ odd, and ${\it O}_{N-1}$ is the dense open subset of $A$'s of maximal
rank.  The orbits are nested:  the boundary
$\overline{ {\it O}_{2k} } - {\it O}_{2k}$ of the orbit ${\it O}_{2k}$ consists
precisely of the union of the ${\it O}_{2i}$ for $i < k$.  For
$k < N-1$, the orbit ${\it O}_{2k}$ is singular along its
boundary.  An easy computation similar to what will be done momentarily 
for $Pf$ shows that the codimension in ${\bf P}(\Lambda^2 V^{\vee})$ of
${\it O}_{N-j}$, for even $N-j$, is $j(j-1)/2$.

Next, let us compute the dimension of the vanishing locus.
This is not a global complete intersection, so we cannot get the dimension
by subtracting the number of equations.  An alternative is to work as follows.
At a generic point on ${\bf P}^6$, the matrix $A$ is rank 6, and so
can be block-diagonalized to the form
\begin{displaymath}
\left[ \begin{array}{c|c|c}
0_3 &  B_3 & 0 \\ \hline
-B_3 & 0_3 & 0 \\ \hline
0 & 0 & 0
\end{array} \right]
\end{displaymath}
where $B_3$ denotes a $3 \times 3$ matrix and
$0_3$ denotes the identically-zero $3 \times 3$ matrix.
Near a subgeneric point, where all the $6 \times 6$ Pfaffians vanish,
so that the matrix has rank 4 at the subgeneric point, the
matrix $A$ can be block-diagonalized to the form
\begin{displaymath}
\left[ \begin{array}{c|c|ccc}
0_2 & B_2 & 0 & 0 & 0 \\ \hline
-B_2 & 0_2 & 0 & 0 & 0 \\ \hline
0 & 0 & 0 & \alpha_1 & \alpha_2 \\
0 & 0 & - \alpha_1 & 0 & \alpha_3 \\
0 & 0 & - \alpha_2 & -\alpha_3 & 0
\end{array} \right]
\end{displaymath}
where the $\alpha_i$ are three functions that become zero on the
rank 4 locus.  Since to get to the locus we must set three functions to
zero, we see that our non-global-complete-intersection must have codimension
three in ${\bf P}^6$, and so must have dimension $6-3 = 3$. 
Similarly, the vanishing locus of both the $6 \times 6$ Pfaffians
as well as the $4 \times 4$ Pfaffians is the locus of matrices of
rank 2, and working locally as above we find that to get to that
locus we would have to locally set $4+3+2+1 = 10$ functions to zero,
and so that locus would be codimension 10.  

Let us now compute
$c_1(T)$ for 
the vanishing locus in ${\bf P}^{m-1}$ of the
$(N-1)\times (N-1)$ Pfaffians of a skew-symmetric $N \times N$
matrix ($N$ odd) linear in the homogeneous coordinates on
${\bf P}^{m-1}$.  (For a more intuitive and less precise way to understand
if a given vanishing locus is Calabi-Yau, see for example 
\cite{hubsch}[section 3.5.2].)

Let $V$ be a vector space of odd dimension $N$.
Define $P = {\bf P}(\Lambda^2 V^{\vee})$, {\it i.e.}, the space of 
alternating two-forms on $V$, modulo scalars.
For a generic $A \in P$, the rank of $A$ is $N-1$.
Now, define 
\begin{displaymath}
Pf \: \equiv \: \{ A \in P \, | \, \mbox{rk }A \leq N-3 \}
\end{displaymath} 
For $A \in Pf$ the kernel is three dimensional.

Now, define 
\begin{displaymath}
\widetilde{Pf} \: \equiv \: \{ (A,s) \, | \, A \in P, \: A s \: = \: 0 \}
\end{displaymath}
This is a partial resolution of $Pf$.
Next, define
\begin{displaymath}
D \: \equiv \: \{ A \in P \, | \, \mbox{rk }A \leq N-5 \}
\: = \: \mbox{singular locus of }Pf 
\end{displaymath}
and
\begin{displaymath}
\tilde{D} \: \equiv \: \{ (A,s) \in \widetilde{Pf} \, | \, 
A \in D \}
\end{displaymath}
The space $\tilde{D}$ is the exceptional divisor in $\widetilde{Pf}$,
which is blown down to the singular locus $D$ in $Pf$.

Pulling back the hyperplane bundles via the two composed maps
\begin{eqnarray*}
\pi: & \widetilde{Pf} & \longrightarrow \: Pf \:
\longrightarrow \: P \: = \: {\bf P}( \Lambda^2 V^{\vee} ) \\
\pi': & \widetilde{Pf} & \longrightarrow \: G(3,V) \: \longrightarrow \:
{\bf P}(\Lambda^3 V)
\end{eqnarray*}
defines line bundles $H$, $H'$, respectively, on $\widetilde{Pf}$.
It can be shown (see below) that the divisor
$\tilde{D}$ can be expressed as a linear combination of $H$, $H'$, with
the result
\begin{equation} \label{ron1}
\tilde{D} \: = \: ((N-3)/2) H \: - \: H'
\end{equation}
on $\widetilde{Pf}$.
Since $\tilde{D}$ is blown down in $Pf$, this linear combination will give
a relation between the corresponding divisor classes on $Pf$:
\begin{equation} \label{ron2}
H' \: = \: ((N-3)/2) H 
\end{equation}
on $Pf$.
One can then calculate $c_1(Pf)$ from its embedding into $P$:
\begin{equation}  \label{ron3}
c_1(Pf) \: = \: c_1(P) \: - \: c_1({\cal N})
\end{equation}
where ${\cal N}$ is the normal bundle to $Pf$ in $P$.
(This is a rank 3 vector bundle away from the singular locus $D$; since $D$
has high codimension, one works away from $D$ to get the correct answer
for $c_1$.)
This normal bundle can be identified as
\begin{equation} \label{ron4}
{\cal N} \: = \: \Lambda^2 S^{\vee} \otimes H
\end{equation}
where $S$ is (the pullback to $Pf - D = \widetilde{Pf} - \tilde{D}$ of)
the universal rank 3 subbundle on $G(3,V)$, fitting into the
exact sequence
\begin{displaymath}
0 \: \longrightarrow \: S \: \longrightarrow \: V \otimes
{\cal O}_{G(3,V)} \: \longrightarrow \: Q \: \longrightarrow \: 0
\end{displaymath}
and as above, $H$ is the pullback of ${\cal O}_P(1)$.
From equation~(\ref{ron4}) we get that
\begin{displaymath}
c_1({\cal N}) \: = \: 2 H' \: + \: 3 H
\end{displaymath}
so from equations~(\ref{ron3}) and (\ref{ron2}) we have
\begin{eqnarray*}
c_1(Pf) & = & ( \dim(P) + 1) H \: - \: 2 H' \: - \: 3H \\
& = & (\dim(P) -2) H - 2 H' \\
& = & ( \dim(P) \: - \: (N-1)) H
\end{eqnarray*}
Finally for
\begin{displaymath}
X \: \equiv \: Pf \mbox{ intersected with a generic linear
subspace } {\bf P}^{m-1}
\end{displaymath}
we get by adjunction
\begin{eqnarray*}
c_1(TX) & = & ( \dim(X) \: - \: (N-4) ) H \\
& = & (m - N) H
\end{eqnarray*}

We still need to justify equation~(\ref{ron1}) in the cohomology
of $\widetilde{Pf}$.  Note that
$\widetilde{Pf}$ is a projective bundle over $G(3,V)$, in fact
it is the projectivization of $\Lambda^2 Q^{\vee}$, where as above,
$Q$ is the universal quotient bundle over $G(3,V)$.  In particular,
it follows that $H^2(\widetilde{Pf})$ is two-dimensional,
so a relation among $H$, $H'$, and $\tilde{D}$ must exist.
We can find the actual relation by intersecting with two 
independent curves $a$, $b$ in $\widetilde{Pf}$.

Define $a$ as follows.  Fix a 3-space $s \in G(3,V)$, and take a generic
pencil of $A$'s in $\Lambda^2((V/s)^{\vee})$, {\it i.e.} which vanish on
this fixed $s$.

Define $b$ as follows.  Take a pencil of $A$'s which does not intersect $D$.

In coordinates $x_1, \cdots, x_N$ on $V$, one can take $a$ to be the line
in $P$ through the points $p$, $q$, and $b$ the line through $p$, $r$, where:
\begin{eqnarray*}
& p: & x_{12} \: + \: x_{34} \: + \: \cdots \: + \: x_{N-4,N-3} \\
& q: & q_{12} x_{12} \: + \: \cdots \: + \: q_{N-4,N-3} x_{N-4,N-3} \\
& r: & x_{23} \: + \: x_{45} \: + \: \cdots \: + \: x_{N-3,N-2}
\end{eqnarray*}
with numerical coefficients $q_{12}, \cdots, q_{N-4, N-3}$ which are
pairwise distinct.  The $A$'s which are linear combinations of
$p$, $q$ vanish on the 3-space $s$ which corresponds to the
last three coordinates ({\it i.e.} is given by the vanishing of
$x_1, \cdots, x_{N-3}$), and generically only there,
but for the $(N-3)/2$ parameter values given by the
$q_{i,i+1}$ the rank drops to $N=5$ and the null space contains
coordinates $i$, $i+1$.  The $A$'s which are linear combinations of
$p$, $r$ have constant rank $N-3$.  The three-dimensional kernel traces
a rational normal curve in Pl\"ucker space, of degree $(N-3)/2$.
It is now straightforward to count intersection points:
\begin{center}
\begin{tabular}{c|cc}
 & $a$ & $b$ \\ \hline
$H$ & $1$ & $1$ \\
$H'$ & $0$ & $(N-3)/2$ \\
$\tilde{D}$ & $(N-3)/2$ & $0$ 
\end{tabular}
\end{center}
This establishes our claim that 
$\tilde{D} = ((N-3)/2)H - H'$, and completes the calculation of
$c_1(TX) = (m-N)H$.

\subsubsection{A complete intersection in a $G(2,5)$ bundle}
\label{g25bdle}

Another analogous example can be built as follows.
Let us consider a Calabi-Yau built as a complete intersection in the
total space of a $G(2,5)$ bundle over ${\bf P}^2$.
The $G(2,5)$ bundle is built by fibering the ${\bf C}^5$ as the
vector bundle ${\cal O}(-1)^{\oplus 4} \oplus {\cal O}$
over ${\bf P}^2$.  The gauged linear sigma model can be built from
the fields $x_0$, $x_1$, $x_2$, $\Phi^{\mu}_i$ ($\mu \in \{1,2\}$,
$i \in \{1, \cdots, 5\}$) in which a $U(2) \times U(1)$ action
has been gauged.  The $U(2)$ acts on the $\Phi_i^{\mu}$ as a set of
five doublets, and leaves the $x_a$ invariant.
The $U(1)$ charges are as follows:
\begin{center}
\begin{tabular}{ccccc} 
$x_0$ & $x_1$ & $x_2$ & $\Phi^{\mu}_{1 \cdots 4}$ & $\Phi^{\mu}_5$ \\ \hline
1 & 1 & 1 & -1 & 0
\end{tabular}
\end{center}
Baryons (Pl\"ucker coordinates) naturally arrange themselves as
\begin{displaymath}
\begin{array}{cl}
\epsilon_{\mu \nu} \Phi^{\mu}_i \Phi^{\nu}_j & i, j \in \{1, \cdots, 4\} \\
\epsilon_{\mu \nu} \Phi^{\mu}_i \Phi^{\nu}_5 & i \in \{1, \cdots, 4\} 
\end{array}
\end{displaymath}
These baryons define an embedding of the total space of this
Grassmannian bundle into the total space of the projective bundle
\begin{displaymath}
{\bf P}\left( {\cal O}(-2)^{\oplus 6} \oplus {\cal O}(-1)^{\oplus 4} 
\right) \: \longrightarrow \: {\bf P}^2
\end{displaymath}
We can build a Calabi-Yau by taking the complete intersection of
five hypersurfaces, so we add to the gauged linear sigma model
five fields $p_i$ of charge $-2$ under $\mbox{det }U(2)$ and
charge $+1$ under the $U(1)$.
The superpotential then has the form
\begin{displaymath}
W \: = \: p_i B^{ijk}(x) \epsilon_{\mu \nu} \Phi^{\mu}_j \Phi^{\nu}_k
\: = \: A^{ij}(x,p) \epsilon_{\mu \nu} \Phi^{\mu}_i \Phi^{\nu}_j
\end{displaymath}
The $5 \times 5$ matrix $A^{ij}$ is skew-symmetric.
For $i, j \leq 4$ it is degree $(1,1)$ in $(x,p)$,
and for $i=5, j \leq 4$ it is degree $(0,1)$ in $(x,p)$.

The analysis of the Landau-Ginzburg point of this model is very
similar to that in the previous section.  It is given by the vanishing
locus of $4 \times 4$ Pfaffians of $A^{ij}$ on ${\bf P}^2 \times {\bf P}^4$,
where the ${\bf P}^4$ arises from the $p_i$ fields.  Briefly, this is
because,
as in \cite{hori1}, on the rank 4 locus, there is a single massless $\Phi$
doublet,
and the vacua run to infinity, whereas on the rank 2 locus
(where the $4\times 4$ Pfaffians vanish), there are three massless
$\Phi$ doublets, and in the IR a single vacuum remains.

The geometry we have found at the Landau-Ginzburg point was discussed
mathematically in \cite{andreithesis}[section 3.2], as an example
of a genus-one-fibered Calabi-Yau threefold, fibered over ${\bf P}^2$,
with an embedding into ${\bf P}^2 \times {\bf P}^4$ of the form described
above.  It is not an elliptic fibration, as it does not have an ordinary
section, but rather merely a 5-section.

The Landau-Ginzburg geometry above, is believed to have the same
derived category of coherent sheaves \cite{andreithesis}
as the complete intersection 
in the $G(2,5)$ bundle\footnote{In \cite{andreithesis}, the dual
was described merely as relative
$\mbox{Pic}^2$ -- the more explicit description as a $G(2,5)$ bundle
was not given there.} 
on ${\bf P}^2$ that sits at the large-radius
point in the gauged linear sigma model above.
Fiberwise \cite{andreipriv}, 
if $G(2,5)$ is cut by a codimension 5 plane, the result is
an elliptic curve, and similarly if the Pfaffian manifold
$Pf(5)$ (which can be understood as $G(2,5)$ again, but describing
2-planes in the dual 5-dimensional vector space) is cut by 
the dual linear space, the result is the same elliptic curve.
The two curves can be identified if one picks a point on each elliptic
curve, or alternatively, the second elliptic curve can be identified
naturally with $\mbox{Pic}^2$ of the first curve.  
When this is done in families, the Grassmannian fibration can be
understood as the relative $\mbox{Pic}^2$ of the Pfaffian fibration,
and it is well-known that such pairs are derived equivalent.

Whether or not these two spaces are birational to one another,
is not known at present.

\subsubsection{A Fano example}

In addition to Calabi-Yau examples, we can also consider
non-Calabi-Yau examples.  In this and subsequent sections, we shall
study several examples of nonabelian GLSM's for Fano manifolds.
These GLSM's will also possess nontrivial Landau-Ginzburg points.
We should note at the beginning, however, that there are some structural
differences between GLSM K\"ahler moduli spaces in Calabi-Yau and
non-Calabi-Yau cases.  For example, in Calabi-Yau cases the
K\"ahler moduli space is complexified by the theta angle,
but in non-Calabi-Yau cases an axial anomaly prevents that theta
angle from being meaningful, so the K\"ahler moduli spaces are real,
not complex.  In addition, the K\"ahler parameters will not be
renormalization-group invariants in non-Calabi-Yau cases.
In the examples below, we will consider some simple
nonabelian GLSM's with a one-real-dimensional K\"ahler moduli ``space.''
Singularities near the origin will break the K\"ahler moduli space into
two disconnected components, and so as there is no way to smoothly
deform the large-radius and Landau-Ginzburg points into one another,
Witten indices need not match.

Let us now consider our first example.
A degree 5 del Pezzo can be described as $G(2,5)[1,1,1,1]$.
At the Landau-Ginzburg point of the GLSM for $G(2,5)[1,1,1,1]$,
following the same analysis as before we find
the vanishing locus in ${\bf P}^3$ of $4 \times 4$ Pfaffians
of a $5 \times 5$ skew-symmetric matrix formed from a linear combination
of the hyperplane equations. 

Note that unlike the Calabi-Yau case, the large-radius and
Landau-Ginzburg points no longer describe spaces of the same
dimension.  The large radius space has complex dimension $6-4 = 2$,
whereas the Landau-Ginzburg phase is codimension three in
${\bf P}^3$ and hence zero-dimension -- a set of
points with multiplicity.

\subsubsection{A related non-Calabi-Yau example}

Consider a GLSM describing the complete intersection of 6 hyperplanes
in $G(2,5)$, {\it i.e.}, $G(2,5)[1,1,1,1,1,1]$
(or, more compactly, $G(2,5)[1^6]$).
The Landau-Ginzburg point of this GLSM can be determined using
previous methods, and is given by the vanishing locus in ${\bf P}^5$
of the $4 \times 4$ Pfaffians of a skew-symmetric $5 \times 5$ matrix
with entries linear in the homogeneous coordinates on 
${\bf P}^5$.

Here, the complete intersection in the Grassmannian $G(2,5)$ has
dimension 0 -- it is a set of points -- whereas the Pfaffian variety
has codimension 3 in ${\bf P}^5$, hence dimension 2, the opposite of the
previous example.

This example is very closely related to the example in the previous
section, as the degree 5 del Pezzo $G(2,5)[1,1,1,1]$ 
(the large-radius point of the previous example) has an alternate
presentation as the vanishing locus in ${\bf P}^5$ of the $4 \times 4$
Pfaffians of a skew-symmetric $5 \times 5$ matrix with entries linear in
the homogeneous coordinates on ${\bf P}^5$, the Landau-Ginzburg point of
the current example.  This is a consequence of the equivalence
described in section~\ref{revpfaff} between $G(2,5)$ and the vanishing
locus in ${\bf P}^9$ of the $4 \times 4$ Pfaffians of a generic
$5 \times 5$ skew-symmetric matrix -- by intersecting both sides
with four hyperplanes, we obtain the equivalence claimed here.

Not only are the dimension-two phases of this example and the previous
one different presentations of the same thing, but so are the dimension-zero
phases.  Both $G(2,5)[1^6]$ and the vanishing locus in ${\bf P}^3$
of the $4\times 4$ Pfaffians of a $5 \times 5$ skew-symmetric matrix
describe the same number of points (including multiplicity).

As a result, we believe the example in this subsection and that 
in the previous subsection are actually different presentations
of the same family of CFT's, in which the interpretation as
large-radius and Landau-Ginzburg points has been reversed.

\subsubsection{More non-Calabi-Yau examples}

Consider a gauged linear sigma model describing the
complete intersection of $m$ hyperplanes with $G(2,N)$ for $N$
odd (using the restriction of \cite{hori1}).
Following reasoning that by now should be routine,
its Landau-Ginzburg point corresponds to the vanishing
locus in ${\bf P}^{m-1}$ of the $(N-1)\times (N-1)$ Pfaffians
of a skew-symmetric $N \times N$ matrix linear in the homogeneous
coordinates on ${\bf P}^{m-1}$, where the $N \times N$ matrix is formed
by rewriting a linear combination of the $m$ hyperplanes, with
coefficients given by the homogeneous coordinates on ${\bf P}^{m-1}$.

The complete intersection in the Grassmannian $G(2,N)$ has dimension
$2(N-2) - m$, whereas the Pfaffian variety has codimension three
in ${\bf P}^{m-1}$, hence dimension $m-4$.
The canonical bundle of the complete intersection is
${\cal O}(m-N)$, whereas following the reasoning in
section~\ref{revpfaff} the canonical bundle of the
Pfaffian variety at the Landau-Ginzburg point is
${\cal O}(-m + N)$. 

The fact that the canonical bundles of either end have
$c_1$'s of opposite sign gives us another check on these
results.  After all, the renormalization of the Fayet-Iliopoulos
parameter is determined by the gauge linear sigma model and is
independent of the phase.  If the large-radius phase 
shrinks under renormalization group flow, {\it i.e.}, 
$r \rightarrow 0$, then to be consistent the theory at Landau-Ginzburg
must expand to larger values of $|r|$, {\it i.e.} $r \rightarrow
- \infty$, and conversely.  A nonlinear sigma model on a positively-curved
space will shrink under RG flow, and a nonlinear sigma model on a
negatively-curved space will expand under RG flow,
so we see that if one phase of the GLSM is positively-curved,
then in order for $dr/d\Lambda$ as determined by the GLSM to be consistent
across both phases, the other phase must be negatively-curved,
and vice-versa.

\subsubsection{Interpretation -- Kuznetsov's homological projective duality}

In the past, it has been thought that the geometric phases of a 
gauged linear sigma model were all birational.  Here, however,
we have seen several examples, both Calabi-Yau and non-Calabi-Yau,
of gauged linear sigma models with non-birational phases.
One natural question to ask then is, if these phases are not
birational, then how precisely are they related mathematically?

In this paper and \cite{alltoappear}, we would like to propose that
these non-birational phases are related by Kuznetsov's ``homological
projective duality'' \cite{kuz1,kuz2,kuz3}.
Put another way, here and in \cite{alltoappear} we propose that
gauged linear sigma models implicitly give a physical realization
of Kuznetsov's homological projective duality.

The physical relevance of Kuznetsov's work will be discussed in
much greater detail in \cite{alltoappear}, together with more examples
(including examples in abelian gauged linear sigma models) and more
tests of corner cases.

Very briefly, two spaces are said to be homologically projectively dual
if their derived categories of coherent sheaves have pieces in common,
in a precise technical sense.  Unfortunately, in general there is not
a constructive method for producing examples of
homological projective duals -- given one space, there is not yet an algorithm
that will always produce a dual.

However, much is known.  For example, it is known that the
examples discussed so far are all homologically projectively dual.
In other words, the complete intersection of $m$ hyperplanes in
$G(2,N)$ for $N$ odd is known to be homologically projectively dual
to the vanishing locus in ${\bf P}^{m-1}$ of the
$(N-1)\times (N-1)$ Pfaffians of a skew-symmetric $N \times N$ matrix
linear in the homogeneous coordinates on ${\bf P}^{m-1}$, where the
$N \times N$ matrix is formed by rewriting a linear combination of
the $m$ hyperplanes.  In special cases, this can be understood more
systematically.  For example, the Grassmannian $G(2,5)$ itself is
homologically projectively dual to a Pfaffian variety of
$5 \times 5$ skew-symmetric matrices, the space of which lies in
${\bf P}(\Lambda^2 {\bf C}^5) = {\bf P}^9$.
Intersecting the Grassmannian with 4 hyperplanes is dual to
intersecting the Pfaffian variety with the dual of ${\bf P}^{9-4}$
in ${\bf P}^9$, which is ${\bf P}^3$.

More generally, we can understand this as follows \cite{kuzpriv310}.
Let $M$ denote the dimension of the set of all skew-symmetric $N \times N$
matrices, $N$ odd,
choose $M$ generic hyperplanes in $G(2,N)$, and consider
the vanishing locus in ${\bf P}^{M-1}$ of the $(N-1)\times (N-1)$ Pfaffians
of a skew-symmetric $N \times N$ matrix linear in the homogeneous
coordinates on ${\bf P}^{M-1}$, where the $N \times N$ matrix is formed by
rewriting a linear combination of the $M$ hyperplanes, with coefficients
given by the homogeneous coordinates on ${\bf P}^{M-1}$.
This variety, call it $Y$, is homologically projectively dual to $G(2,N)$.
If we are given $m$ hyperplanes in $G(2,N)$, they generate 
${\bf P}^{m-1} \subset {\bf P}^{M-1}$, and so the dual is
$Y \cap P^{m-1}$.
In the language of section~\ref{revpfaff}, this duality exchanges
orbits ${\it O}_2 \leftrightarrow {\it O}_{N-3}$, as well
as intersections:  ${\it O}_2$ intersected with a codimension $m$ subspace
$A$ is dual to ${\it O}_{N-3}$ intersected with $A^{\perp}$, which is
a ${\bf P}^{m-1}$.
The general result verifies not only the previous example
but every example discussed so far, except for the $G(2,5)$ bundle model
in section~\ref{g25bdle}.
For that model, it is believed \cite{kuzpriv0403} that the 
two geometries are homologically projectively dual, though a rigorous
proof does not yet exist.

Strictly speaking, what often arises as the duals are not precisely
ordinary nonlinear sigma models on spaces, but rather
certain resolutions known technically as `noncommutative resolutions.'
The term `noncommutative' is a misnomer here, as it does not necessarily
imply that there is any noncommutative algebra present, but rather is
used as a generic term for anything on which one can do sheaf theory,
and `noncommutative spaces' are defined by their sheaves.
For example, via matrix factorization one can think of a 
Landau-Ginzburg model as a `noncommutative space' in this sense.
In \cite{alltoappear} we will closely examine some examples of
homological projective duality in which noncommutative resolutions
appear, and we will see how those noncommutative resolutions arise
within gauged linear sigma models.

So far we have outlined how non-birational phases of gauged linear sigma
models can be understood as examples of Kuznetsov's homological projective
duality, but we also conjecture the same is true of birational phases
also.  It is known that some flops between Calabi-Yau's are examples
of homological projective duals, and it is conjectured that all flops
are also examples of homological projective duals.

However, although evidence suggests that all phases of gauged linear
sigma models might be examples of homological projective duality,
the converse may not be true -- there are examples of homological projective
duals which might not be realizable within gauged linear sigma models.
For example, Grassmannians themselves have homological projective duals,
but a typical gauged linear sigma model for a Grassmannian has only
one phase.  Perhaps there are alternate presentations, realizable
within gauged linear sigma models, which have multiple phases,
or perhaps the duals reflect more subtle aspects of the original
gauged linear sigma models.  

In the next subsection, we shall consider a possible prediction of
Kuznetsov's work for cases in which the physics is more obscure.

\subsubsection{Duals of $G(2,N)$ with $N$ even}

The authors of \cite{hori1} restricted to their analysis of duality in complete
intersections in $G(2,N)$ to the case $N$ odd,
because the physics in cases for which $N$ is even is both different
and more difficult.
However, there exist homological projective duals for cases in
which $N$ is even, and so we can use homological projective duality
to make predictions for such cases.    

The duals of $N$ even cases are slightly subtle,
so we must proceed carefully.
Let us first examine the case of $G(2,6)[1^6]$ in detail.
Naively,
the homological projective dual of the complete
intersection $G(2,6)[1^6]$ is a `noncommutative resolution'
of a cubic 4-fold, a hypersurface in ${\bf P}^5$ defined by the
Pfaffian of a skew-symmetric $6 \times 6$ matrix determined by the
six hyperplanes in $G(2,6)$.

Given the match between homological projective duals and phases in
previous examples, it is tempting to conjecture that the Landau-Ginzburg
phases of $G(2,6)[1^6]$ might be the Pfaffian variety above.
On the other hand, although the complete intersection is Calabi-Yau,
the Pfaffian variety is not.  It is difficult to understand how
a gauged linear sigma model could relate a Calabi-Yau space to a
non-Calabi-Yau space, unless the non-Calabi-Yau space has a flux
background present so as to preserve supersymmetry.

In fact, the correct answer here is slightly more subtle.
The correct dual \cite{kuz1,tpantevpriv} is a
`noncommutative\footnote{As described earlier,
but is well worth repeating, the word `noncommutative'
is used by mathematicians in this context in a misleading fashion -- it
does not necessarily indicate the presence of any noncommutivity.}
space' 
defined by a subcategory of the derived
category of the cubic.  Two spaces are dual in the relevant
mathematical sense if, in essence, their derived categories have
a piece in common \cite{kuz3}[theorem 2.9].  For $G(2,N)[1^N]$ for $N$ odd,
the corresponding complete intersection in the Pfaffian variety is
already a Calabi-Yau.  
The derived category of the complete intersection in the Pfaffian
variety already matched\footnote{We should be slightly careful.
For $N$ odd and $N>7$, there is a technicality involving the
construction of a (noncommutative) resolution of the Pfaffian,
so to be precise, these remarks ought to be limited to small $N$,
and considered to be conjectures for larger $N$.} 
that of the Calabi-Yau $G(2,N)[1^N]$
(see {\it e.g.} \cite{kuz3,andreilev}),
so there was no `noncommutative space' to consider.
However, for $G(2,N)[1^N]$ for
$N$ even, the derived category of the original space matches only a 
subcategory of that of the hypersurface in the Pfaffian.
So the correct dual in the case of $G(2,6)[1^6]$  
must be something slightly different from the cubic 4-fold described
above, something defined by a subcategory of the derived category of
the cubic 4-fold.  We then have to find a physical theory whose
D-branes are described by the pertinent subcategory.

It turns out that there is a very natural candidate for such a physical
theory, that should realize the `noncommutative space' defined by a 
subcategory of the derived category of the cubic 4-fold.
In particular, since the cubic is Fano, the category of matrix
factorizations of the associated Landau-Ginzburg model is a subcategory
\cite{orlov1}, and is precisely the subcategory we need to describe.  
That Landau-Ginzburg model is defined by a cubic
superpotential in six variables, which up to finite group factors
is a deformation of a Landau-Ginzburg model for an orbifold of 
a product of two elliptic curves -- a K3, in other words.
Assuming that the finite group factors work out correctly,
and that we have interpreted the structure of that model and its
deformations correctly,
this finally gives us a physically consistent conjecture for the
form of the LG point in $G(2,6)[1^6]$ -- namely, that it is a Landau-Ginzburg
model associated to a K3 surface.

So, based on the mathematics of Kuznetsov's homological projective
duality, we propose (as suggested to us by
\cite{tpantevpriv}) that the Landau-Ginzburg point of the
nonabelian GLSM for $G(2,6)[1^6]$ is a Landau-Ginzburg model corresponding
to a K3 surface.

We will leave to future work a thorough verification that this is indeed
the correct description of the Landau-Ginzburg point of this model,
although we will mention that it seems very plausible:
\begin{itemize}
\item As a consistency check, since $G(2,6)[1^6]$ is itself
a K3, the only other geometries one should find as phases in a 
GLSM K\"ahler moduli space are other K3's, so the result is physically
consistent.
\item At the level of sheaf theory and D-branes \cite{medercat,mikedc},
both the derived category of $G(2,6)[1^6]$ and the category of
matrix factorizations of the Landau-Ginzburg model appear 
\cite{tpantevpriv} as subcategories
of the derived category of the cubic fourfold in ${\bf P}^5$ defined
as the right orthogonal to ${\cal O}(k)$ for $k=-1,-2,-3$, 
{\it i.e.}, ${\cal F}$ such that ${\bf R}\mbox{Hom}({\cal O}(k), 
{\cal F}) = 0$ for $k=-1,-2,-3$.  So the derived categories of the
large-radius limit and the proposed interpretation of the
Landau-Ginzburg point match on the nose, exactly as needed for consistency
of the B twist of the gauged linear sigma model.
\item Physically, this would mean that the fields fluctuate around
a single vacuum at energy cost determined by the Pfaffian of
a skew-symmetric $6 \times 6$ matrix, which is very plausibly
a description of the Landau-Ginzburg point of this gauged linear sigma
model.
\end{itemize}

In addition, we should also mention one other, related, geometry that
appears in this context.  It is less clear to us what role it plays
physically, but for completeness, it should be mentioned.
The space of lines in the cubic in ${\bf P}^5$ is another Calabi-Yau -- 
in fact, a hyperK\"ahler 4-fold given by the Hilbert scheme of pairs of
points on the K3 $G(2,6)[1^6]$.  Its derived category is nearly 
(but not quite) identical to that of $G(2,6)[1^6]$.
For more information on this space and its relation
to $G(2,6)[1^6]$ see for example \cite{beauvilledonagi}.

Similarly, the dual of $G(2,8)[1^8]$ is a Landau-Ginzburg model
defining a `noncommutative space' associated to a 
`noncommutative resolution' of a degree four hypersurface in ${\bf P}^7$,
where the hypersurface is defined by the Pfaffian of a skew-symmetric
$8 \times 8$ matrix formed from the eight hyperplanes.
The associated Landau-Ginzburg model is
defined by a degree four superpotential in eight variables,
which is related to an orbifold of the product of two K3 surfaces.
In this case we conjecture that the Landau-Ginzburg point of the
GLSM for $G(2,8)[1^8]$ is an orbifold of a Landau-Ginzburg model with
a quartic superpotential in eight variables, associated to
$(K3 \times K3)/G$ for some $G$.

Technically \cite{kuzpriv310}, we can understand the duals as follows.
The complete intersection in a Pfaffian $Y$ dual to $G(2,N)$ for $N$ even is
(a certain resolution of) the set of all degenerate skew-symmetric matrices
in ${\bf P}^{M-1}$, where $M$ is the dimension of the set of all
skew-symmetric $N \times N$ matrices.  Such $Y$ can be described as a 
hypersurface in ${\bf P}^{M-1}$ of degree $N/2$, defined by the
vanishing locus of the Pfaffian.  The dual to a complete
intersection of $m$ hyperplanes in $G(2,N)$ is the intersection 
$Y \cap {\bf P}^{m-1}$.
Thus, for example, the variety dual to $G(2,6)[1^6]$ is a degree $6/2=3$
hypersurface in ${\bf P}^{6-1} = {\bf P}^5$.

However, these spaces are never Calabi-Yau, and so can not be
the physically-realized homological projective duals to the
Calabi-Yau's $G(2,N)[1^N]$.  
More generally \cite{kuzpriv310},
a Calabi-Yau built as a complete intersection of
$N$ hyperplanes in $G(2,N)$ odd will always have a Calabi-Yau
homological projective dual; but for $N$ even, the
dual variety obtained as a complete intersection in a Pfaffian
variety will never be Calabi-Yau (leading us to work with related
Landau-Ginzburg models).

Repeating the analysis above, we conjecture that the Landau-Ginzburg points
of GLSM's for $G(2,N)[1^N]$ for $N$ even will always be the
`correct' homological projective duals, namely Landau-Ginzburg
models defined by degree $N/2$ superpotentials given by the Pfaffian
of the skew-symmetric $N\times N$ matrix defined by the hyperplanes.
The category of matrix factorizations of these Landau-Ginzburg models
will match the derived category of $G(2,N)[1^N]$ for all even $N$,
as needed for consistency of the B twist of the gauged linear sigma
model, and will also be a large distinguished subcategory of
the derived category of the pertinent degree $N/2$ hypersurface in
${\bf P}^{N-1}$.

As an aside, the reader might wonder whether $G(2,N)[1^N]$ is hyperK\"ahler for
all even $N$, since our duals above both at least naively appear
hyperK\"ahler, and $G(2,N)[1^N]$ always has even complex dimension.
However, it can be shown that $G(2,N)[1^N]$ cannot be hyperK\"ahler for
$N > 6$.  This is because, due to Lefschetz, $h^{2,0}$ of
the complete intersection $G(2,N)[1^N]$ matches that of the
ambient space $G(2,N)$, which vanishes, but on the other hand,
hyperK\"ahler implies holomorphically symplectic which implies
$h^{2,0} = 1$.

\subsection{Non-birational derived equivalences in abelian GLSMs}

Although the example in section~\ref{horitongrodland}
is sometimes advertised as the
first example of a gauged linear sigma model involving
(a) a Calabi-Yau not presented as a complete intersection, and
(b) two non-birationally-equivalent Calabi-Yau's, in fact an
abelian gauged linear sigma model with both these properties
appeared previously in \cite{ps4}[section 12.2].
There, the gauged linear sigma model described a complete intersection
of four degree two hypersurfaces in ${\bf P}^7$ for $r \gg 0$,
and was associated with a double cover of ${\bf P}^3$ branched
over a degree eight hypersurface (Clemens' octic double solid)
for $r \ll 0$.  The fact that they are not birational follows from
the fact that \cite{grosspriv} the complete intersection in
${\bf P}^7$ has no contractible curves, whereas the branched double
cover has several ordinary double points.
This is another example of Kuznetsov's homological projective
duality \cite{kuz2,alltoappear}.

The double cover structure at the Landau-Ginzburg limit arises because
generically the only massless fields have nonminimal charges.
After all, the superpotential
\begin{displaymath}
W \: = \: \sum_i p_i Q_i(x)
\end{displaymath}
(where the $Q_i$ are quadric polynomials)
can be equivalently rewritten in the form
\begin{displaymath}
W \: = \: \sum_{ij} x_i A^{ij}(p) x_j
\end{displaymath}
where $A^{ij}$ is a symmetric matrix with entries linear in the $p$'s.
Away from the locus where $A$ drops rank, {\it i.e.}, away from the
hypersurface $\mbox{det }A = 0$, the $x_i$ are all massive,
leaving only the $p_i$ massless, which all have charge $-2$.
A GLSM with nonminimal charges describes a gerbe \cite{ps1,ps2,stxglsm},
and physically a string on a gerbe is isomorphic to a string on a disjoint
union of spaces \cite{ps4} (see \cite{ps5} for a short review).
Thus, away from the branch locus, we have a double cover,
and there is a Berry phase (see \cite{alltoappear}) that interweaves
the elements of the cover, thus giving precisely the right global
structure.

Mathematically, the branched double cover will be singular
when the branch locus is singular.
(More generally, if the branch locus looks like $f(x_1, \cdots, x_n) = 0$,
then the
double cover is given by $y^2=f(x_1,\cdots,x_n)$, and it is straightforward
to check that the double cover $y^2=f$ will be smooth precisely where
the branch locus $f=0$ is smooth.)
Physically, however, there is no singularity in these models.
After all, the $F$ term conditions in this model can be written
\begin{eqnarray*}
\sum_j A^{ij}(p) x_j & = & 0 \\
\sum_{ij} x_i \frac{ \partial A^{ij} }{ \partial p_k } x_j & = &
0
\end{eqnarray*}
and it can be shown that these equations have no solution -- the GLSM
is smooth where the geometry is singular.
This is one indication that the Landau-Ginzburg point of this GLSM
is actually describing a certain (`noncommutative') resolution of the
branched double cover, matching the prediction of Kuznetsov's homological
projective duality, as will be described in greater detail in
\cite{alltoappear}.

Mathematically, the double cover can be understood
as a moduli space of certain bundles on the complete intersection
of quadrics \cite{grosspriv}.  (Each quadric in ${\bf P}^7$ carries two
distinct spinor bundles which restrict to bundles on the complete
intersection, and when the quadric degenerates, the spinor bundles
become isomorphic, hence giving the double cover of ${\bf P}^3$.)

Also, the twisted derived category of coherent sheaves of the
double cover is expected to be
\cite{grosspriv} isomorphic to the derived category of the
complete intersection; such a derived equivalence is to be expected
for two geometries on the same GLSM K\"ahler moduli space.

The primary physical difference between the gauged linear sigma model
in \cite{hori1}[section 5], reviewed in section~\ref{horitongrodland},
and \cite{ps4}[section 12.2], reviewed here, 
is that in the
former, at least one $\phi$ always remains massless (and is removed by quantum
corrections), whereas in the latter all of the $\phi$ are generically
massive.  Thus, in the latter case one generically has a nonminimally
charged field, $p$, and so gerbes are relevant, whereas in the
former there is never a nonminimally-charged-field story.

Also, examples of this form generalize easily.
The complete intersection of $n$ quadrics in ${\bf P}^{2n-1}$ is
related, in the same fashion as above, to a branched double cover
of ${\bf P}^{n-1}$, branched over a determinantal hypersurface
of degree $2n$.  These are Calabi-Yau, for the same reasons as
discussed in \cite{ps4}[section 12.2].
(In fact, when $n=2$, the branched double cover is just the
well-known expression of certain elliptic curves as branched double
covers of ${\bf P}^1$, branched over a degree four locus.
The fact that K3's can be described as double covers branched over
sextic curves, as realized here for $n=3$,
is described in \cite{gh}[section 4.5], and the relation between
the branched double cover and the complete intersection of
quadrics is discussed in \cite{mukai1}[p. 145].)
This mathematical relationship can be
described physically with gauged linear sigma models of
essentially the same form as above.

These examples, and Kuznetsov's homological projective duality,
will be discussed in greater detail in \cite{alltoappear}.

We should also point out that examples of this form indicate
that gerbe structures
are much more common in GLSM's than previously thought.
At minimum, we see that analysis of GLSM's requires understanding
the role stacks play in physics.  Stacks were originally introduced
to physics
both to form potentially new string compactifications, and to better
understand physical characteristics of string orbifolds, which are
described by special cases of strings on stacks.
The initial drawbacks included the fact that string compactifications
on stacks seemed to suffer from physical inconsistencies,
including deformation theory inconsistencies and manifest violations of
cluster decomposition, which might be a signal of a potentially
fatal presentation dependence.  These issues were explored
and resolved in \cite{stxglsm,ps4,ps1,ps2,ps5}.
In some sense, this paper and \cite{alltoappear} represent a continuation
of that work.

\section{Quantum cohomology}
\label{quantcohom}

In this section we collect some physical analyses pertinent
to quantum cohomology rings of Grassmannians and flag manifolds.

We begin in subsection~\ref{qcbasics}
with a brief review of known results for quantum
cohomology rings of such spaces.  
Quantum cohomology of flag manifolds has been studied in a variety
of references; see for 
example\footnote{This is merely a sample list of references -- it is not 
intended
to be complete or comprehensive.} 
\cite{witver,bdw,sadov1,cf1,cf2,cf3,kim1,kim2,bck0,bck,lchen1,stromme,llly}.
In particular, \cite{sadov1} contains a more nearly physics-oriented
discussion of nonlinear sigma models.

In subsection~\ref{qc1loop} we
briefly outline how those results can be understood in terms of
one-loop effective actions in nonabelian gauged linear sigma models.

Finally, in subsection~\ref{qclsm},
we discuss `linear sigma model' moduli spaces
of maps from ${\bf P}^1$ into Grassmannians and flag manifolds.  These spaces,
which are compactifications of spaces of maps,
figure prominently in physics discussions of quantum cohomology
rings and related invariants.  Although they are well-understood
for toric varieties, they are less well understood for 
Grassmannians and flag manifolds.  More specifically,
there is a natural mathematical notion of what those spaces should be
-- namely Quot and hyperquot schemes -- and those mathematical notions
have often been used in mathematical treatments of quantum cohomology
rings of Grassmannians and flag manifolds.
However, unlike the case of toric varieties, it has not been
clear how to get those spaces from thinking about nonabelian
gauged linear sigma models, and in fact it has not been completely
obvious that the physically relevant `linear sigma model' moduli spaces
really are the same as Quot's and hyperquot's, rather than
some spaces birationally equivalent to them.  
In the last section we will perform the physical
computation of the LSM moduli spaces in several examples, and demonstrate
explicitly that physics really does see precisely the same
Quot's and hyperquot's that are mathematically relevant.

\subsection{Basics}   \label{qcbasics}

In this subsection we shall briefly review known results for
quantum cohomology rings for Grassmannians and flag manifolds.

The quantum cohomology ring of the Grassmannian $G(k,N)$ can be
very compactly expressed as follows \cite{witver}[section 3.2].
Let $S$, $Q$ be the universal subbundle, quotient bundle as described
earlier, then the Chern classes of $S$ generate the cohomology of 
$G(k,N)$, and the sole modification of quantum corrections is to
change the classical constraint\footnote{Given the short exact sequence
relating $S$ and $Q$, the total Chern classes are necessarily related
by $c(S^{\vee})c(Q^{\vee}) = 1$.  The first $N-k$ cohomology classes
in the expansion can be used to solve for the Chern classes of
$Q^{\vee}$ in terms of those of $S^{\vee}$; the remaining ones are
relations among the cohomology classes, including the one below at the
last step.}
\begin{displaymath}
c_k(S^{\vee}) c_{N-k}(Q^{\vee}) \: = \: 0
\end{displaymath}
to the constraint 
\begin{equation}  \label{grassqcoh}
c_k(S^{\vee}) c_{N-k}(Q^{\vee}) \: =\: q
\end{equation}

It is straightforward to check that the result above reproduces
the well-known quantum cohomology ring of a projective space.
Recall ${\bf P}^{N-1} = G(1,N)$, so in terms of the Grassmannian we have
the constraint
\begin{displaymath}
c_1(S^{\vee}) c_{N-1}(Q^{\vee}) \: = \: 1
\end{displaymath}
Since the tangent bundle is $S^{\vee} \otimes Q$ and $S$ is a line
bundle, we have that $Q^{\vee} = S^{\vee} \otimes T^* {\bf P}^{N-1}$.
Let $J$ denote the degree-two generator of the cohomology ring of
${\bf P}^{N-1}$.  Then
\begin{eqnarray*}
c_{N-1}( Q^{\vee} ) & = & c_{N-1}(T^* {\bf P}^{N-1}) \: + \:
c_1(S^{\vee}) c_{N-2}(T^* {\bf P}^{N-1}) \: + \:
c_1(S^{\vee})^2 c_{N-3}(T^* {\bf P}^{N-1}) \: + \\
& & \:
\cdots \: +
c_1(S^{\vee})^{N-1} \\
& = & (-)^{N-1}\left( \begin{array}{c} N \\ N-1 \end{array} \right) J^{N-1}
\: + \: (-)^{N-2}\left( \begin{array}{c} N \\ N-2 \end{array} \right) J^{N-1}
\: + \: \cdots \: + \:
\left( \begin{array}{c} N \\ 0 \end{array} \right) J^{N-1} \\
& = & (-)^{N-1} J^{N-1}
\end{eqnarray*}
where we have used the fact that $S^{\vee} \cong {\cal O}(1)$
and $c_i(T^* {\bf P}^{N-1}) = (-)^i \left( \begin{array}{c}
N \\ i \end{array} \right)$.
Thus, the quantum relation
\begin{displaymath}
c_k(S^{\vee}) c_{N-k}(Q^{\vee}) \: =\: q
\end{displaymath}
reduces in the special case $k=1$ to the relation
\begin{displaymath}
J^{N} \: = \: (-)^{N-1} q
\end{displaymath}
which up to irrelevant signs is the quantum cohomology relation on
${\bf P}^{N-1}$.

For the flag manifold $F(k_1, \cdots, k_n, N)$, the quantum cohomology
is somewhat more complicated to express \cite{sadov1}[section 5].
The classical cohomology ring is generated by the Chern classes
of the bundles $S_i/S_{i-1}$.
Let $x^{(i)}_j$ denote the $j$th Chern class of $S_i/S_{i-1}$
(in conventions where $S_0 = 0$ and $S_{n+1} = {\cal O}^{\oplus N}$), 
then the classical cohomology of the
flag manifold can be expressed as the ring
\begin{displaymath}
{\bf C}[x^{(1)}_1, \cdots, x^{(1)}_{k_1}, 
x^{(2)}_{1}, \cdots, x_{N - k_n}^{(n+1)}]
\end{displaymath}
modulo the ideal generated by the coefficients of the polynomial
\begin{displaymath}
P(\lambda) \: = \: \lambda^N \: - \: \prod_{i=1}^{n+1}\left(
\lambda^{k_i - k_{i-1}} \: + \: \lambda^{k_i - k_{i-1} - 1}x_1^{(i)}
\: + \: \cdots \: + \: x_{k_i - k_{i-1}}^{(i)} \right)
\end{displaymath}

Let us check that this reproduces the classical cohomology
ring of the Grassmannian $G(k,N)$.
Following the procedure above, the cohomology is given by the ring
\begin{displaymath}
{\bf C}[x^{(1)}_1, \cdots, x^{(1)}_k, x^{(2)}_1, \cdots,
x^{(2)}_{N-k}]
\end{displaymath}
modulo the ideal generated by coefficients of the polynomial above,
which in this case becomes
\begin{equation}  \label{classreln}
\lambda^N \: - \: \left( \lambda^k \: + \:
\lambda^{k-1} x^{(1)}_1 \: + \: \cdots \: + \: x^{(1)}_k \right)
\left( \lambda^{N-k} \: + \: \lambda^{N-k-1} x^{(2)}_1 \: + \:
\cdots \: + \: x^{(2)}_{N-k}\right)
\end{equation}
Multiplying this out, we find that the coefficient of
$\lambda^N$ is $0$, the coefficient of $\lambda^{N-1}$ is
\begin{displaymath}
x^{(1)}_1 \: + \: x^{(2)}_1
\end{displaymath}
the coefficient of $\lambda^{N-2}$ is
\begin{displaymath}
x^{(1)}_2 \: + \: x^{(1)}_1 x^{(2)}_1 \: + \: x^{(2)}_2
\end{displaymath}
and so forth.
Now, let us compare to the Grassmannian.
The $x^{(i)}_j$ are the Chern classes of the universal subbundle $S$ and
quotient bundle $Q$:
\begin{displaymath}
x^{(1)}_i \: = \: c_i(S), \: \: \:
x^{(2)}_i \: = \: c_i(V/S) \: = \: c_i(Q)
\end{displaymath}
Because $S$ and $Q$ are related by the short exact sequence
\begin{displaymath}
0 \: \longrightarrow \: S \: \longrightarrow \: V \:
\longrightarrow \: Q \: \longrightarrow \: 0
\end{displaymath}
we have that $c(S) c(Q) = 1$, where $c$ denotes the total 
Chern class.  But 
\begin{displaymath}
c(S) c(Q) \: = \: 1 \: + \: \left( c_1(S) + c_1(Q) \right) \: + \:
\left( c_2(S) + c_1(S) c_1(Q) + c_2(Q) \right) \: + \: \cdots
\end{displaymath}
and so we see the coefficients of $c(S) c(Q)$ at any given order
in cohomology are the same as the coefficients of the polynomial
$P(\lambda)$ above.  Thus, we recover the classical cohomology ring
of the Grassmannian as a special case of the classical cohomology ring
of the flag manifold above.

The quantum cohomology of the flag manifold is obtained by deforming the
relations as follows \cite{sadov1}.
Introduce complex parameters $q_1, \cdots, q_n$, and instead of
taking the relations to be defined by the coefficients of $\lambda$
in the polynomial $P(\lambda)$ above, take them to be the coefficients
of $\lambda$ in the polynomial
\begin{displaymath}
\lambda^N \: - \: \det( A + \lambda I )
\end{displaymath}
where $A$ is the $N\times N$ matrix defined by
\begin{displaymath}
\left(
\begin{array}{cccccccccccc}
x_1^{(1)} & \cdots & x^{(1)}_{k_1} & 0 & \cdots &
- (-)^{k_2-k_1} q_1 & \cdots & 0 & \cdots & \cdots & 0 & 0 \\
-1 & \cdots & 0 & 0 & \cdots & 0 & \cdots & 0 & \cdots & \cdots & 0 & 0 \\
\vdots & \ddots & \vdots & \vdots & \ddots & \vdots & \ddots & \vdots &
\ddots & \ddots & \vdots & \vdots \\
0 & \cdots & -1 & x_1^{(2)} & \cdots & x^{(2)}_{k_2-k_1} & \cdots & 
-(-)^{k_3-k_2} q_2 & \cdots & \cdots & 0 & 0 \\
\vdots & \ddots & \vdots & \vdots & \ddots & \vdots & \ddots &  &
\ddots & \ddots & \vdots & \vdots \\
\vdots & \ddots & \vdots & \vdots & \ddots & \vdots & \ddots &  &
\ddots & \ddots & \vdots & \vdots \\
0 & \cdots & 0 & 0 & \cdots & 0 & \cdots & \cdots & x_1^{(n+1)} & \cdots &
x^{(n+1)}_{N-k_n-1} & x^{(n+1)}_{N-k_n} \\
0 & \cdots & 0 & 0 & \cdots & 0 & \cdots & \cdots & -1 & \cdots & 0 & 0 \\
\vdots & \ddots & \vdots & \vdots & \ddots & \vdots & \ddots & \ddots &
\vdots & \ddots & \vdots & \vdots \\
0 & \cdots & 0 & 0 & \cdots & 0 & \cdots & \cdots & 0 & \cdots & -1 & 0
\end{array}
\right)
\end{displaymath}

In the special case of the Grassmannian $G(k,N)$,
the matrix $A + \lambda I$ is given by
\begin{displaymath}
\left(
\begin{array}{cccccccccc}
x^{(1)}_1 + \lambda & x^{(1)}_2 & \cdots & x^{(1)}_{k-1} & x^{(1)}_k & 0 & 0
& \cdots & 0 & -(-)^{N-k}q \\
-1 & \lambda & \cdots & 0 & 0 & 0 & 0 & \cdots & 0 & 0 \\
\vdots & \ddots & \ddots & \vdots & \vdots & \vdots & \vdots & \cdots & 
\vdots & \vdots \\
0 & 0 & \ddots & \lambda & 0 & 0 & 0 & \cdots & 0 & 0 \\
0 & 0 & \cdots & -1 & \lambda & 0 & 0 & \cdots & 0 & 0 \\
0 & 0 & \cdots & 0 & -1 & x^{(2)}_1 + \lambda & x^{(2)}_2 & \cdots
& x^{(2)}_{N-k-1} & x^{(2)}_{N-k} \\
0 & 0 & \cdots & 0 & 0 & -1 & \lambda & \cdots & 0 & 0 \\
0 & 0 & \cdots & 0 & 0 & 0 & -1 & \cdots & 0 & 0 \\
\vdots & \vdots & \cdots & \vdots & \vdots & \vdots & \vdots & \ddots & 
\vdots & \vdots \\
0 & 0 & \cdots & 0 & 0 & 0 & 0 & \cdots & \lambda & 0 \\
0 & 0 & \cdots & 0 & 0 & 0 & 0 & \cdots & -1 & \lambda
\end{array} \right)
\end{displaymath}
It can be shown that
\begin{eqnarray*}
\lefteqn{ \det(A + \lambda I) \: = \: \hspace{3in} } \\
& & \left( \lambda^k \: + \:
\lambda^{k-1} x^{(1)}_1 \: + \: \cdots \: + \: x^{(1)}_k \right)
\left( \lambda^{N-k} \: + \: \lambda^{N-k-1} x^{(2)}_1 \: + \:
\cdots \: + \: x^{(2)}_{N-k}\right)
\: + \: (-)^{N-k-1}q
\end{eqnarray*}
The relations in the polynomial ring are the coefficients of
\begin{displaymath}
\lambda^N \: - \: \det(A + \lambda I)
\end{displaymath}
Comparing to equation~(\ref{classreln}) we see that the relations for
the quantum cohomology ring are nearly identical to those for
the classical cohomology ring, except for the
one modification
\begin{displaymath}
x^{(1)}_k x^{(2)}_{N-k} \: + \: (-)^{N-k-1} q \: = \: 0
\end{displaymath}
which the reader will recognize as the
quantum cohomology relation for the Grassmannian described
earlier in equation~(\ref{grassqcoh}), up to an irrelevant sign.

For the flag manifold $F(k_1, k_2, N)$, it can be shown that
\begin{displaymath}
\det(A + \lambda I) \: = \: Q_1 Q_2 Q_3 \: + \: (-)^{N-k_2-1} q_2 Q_1
\: - \: (-)^{k_2-k_1-1} q_1 Q_3
\end{displaymath}
where
\begin{eqnarray*}
Q_1 & = & \lambda^{k_1} \: + \: \lambda^{k_1-1} x^{(1)}_1 \: + \: \cdots 
\: + \: x^{(1)}_{k_1} \\
Q_2 & = & \lambda^{k_2 - k_1} \: + \: \lambda^{k_2-k_1-1} x^{(2)}_1 
\: + \: \cdots \: + \:
x^{(2)}_{k_2-k_1} \\
Q_3 & = & \lambda^{N-k_2} \: + \: \lambda^{N-k_2-1} x^{(3)}_1 
\: + \: \cdots \: + \:
x^{(3)}_{N-k_2}
\end{eqnarray*}
As before, when $q_1=q_2=0$, the quantum cohomology relations one
derives from the above reduce to the classical cohomology relations.
For cohomology degrees lower than both $k_2$ and $N-k_1$, the 
relations one derives from the above are identical to the classical
relations.  However, in lower degrees there are quantum corrections.
From the coefficient of $\lambda^{N-k_2}$ we get the relation
\begin{eqnarray*}
\lefteqn{ x^{(1)}_{k_1} x^{(2)}_{k_2-k_1} \: + \:
x^{(1)}_{k_1-1} x^{(2)}_{k_2-k_1} x^{(3)}_1 \: + \:
x^{(1)}_{k_1} x^{(2)}_{k_2-k_1-1} x^{(3)}_1 \: + \: \cdots
} \\
& & \: + \: 
(-)^{k_2-k_1} q_1 \: + \: (-)^{N-k_2-1} q_2 x^{(1)}_{k_1+k_2-N} \: = \: 0
\end{eqnarray*}
From the coefficient of $\lambda^{N-k_2-1}$ we get the relation
\begin{eqnarray*}
\lefteqn{ x^{(1)}_{k_1} x^{(2)}_{k_2-k_1} x^{(3)}_1 \: + \:
x^{(1)}_{k_1-1} x^{(2)}_{k_2-k_1} x^{(3)}_2 \: + \:
x^{(1)}_{k_1} x^{(2)}_{k_2-k_1-1} x^{(3)}_2 \: + \: \cdots } \\
& & \: + \:
(-)^{k_2-k_1} q_1 x^{(3)}_1 \: + \:
(-)^{N-k_2-1} q_2 x^{(1)}_{k_1+k_2+1-N} \: = \: 0
\end{eqnarray*}
From the coefficient of $\lambda^{k_1}$ we get the relation
\begin{eqnarray*}
\lefteqn{ x^{(2)}_{k_2-k_1} x^{(3)}_{N-k_2} \: + \:
x^{(1)}_1 x^{(2)}_{k_2-k_1-1} x^{(3)}_{N-k_2} \: + \:
x^{(1)}_1 x^{(2)}_{k_2-k_1} x^{(3)}_{N-k_2-1} \: + \: \cdots } \\
& & \: + \:
(-)^{N-k_2-1} q_2 \: + \: (-)^{k_2-k_1} q_1 x^{(3)}_{N-k_2-k_1}
\:  = \: 0
\end{eqnarray*}
and so forth.
In each case, $q_1$ acts like a cohomology class of degree $k_2$,
and $q_2$ acts like a cohomology class of degree $N-k_1$.

\subsection{One-loop effective action arguments}  \label{qc1loop}

In \cite{mp}, Batyrev's conjecture for quantum cohomology rings
of toric varieties was derived physically
from the one-loop effective
action for the two-dimensional gauged linear sigma model.
Those techniques were extended to toric stacks in \cite{stxglsm},
and in this subsection we shall briefly outline some heuristics showing how
analogous results arise in 
gauged linear sigma models for flag manifolds.
We will not claim to have a derivation, but rather merely hope to
shed some insight into how the quantum cohomology ring of a flag manifold
arises at the level of nonabelian gauged linear sigma models.

First, recall from \cite{witver}[section 4.2] that in the gauged linear
sigma model for a Grassmannian, the field $\sigma$ is identified in the
low-energy theory with the curvature form for the $U(k)$ gauge
symmetry, which mathematically generates the cohomology of $G(k,N)$.
One way to see this identification of $\sigma$ involves the
equations of motion for $\sigma$ at low energies.  The relevant part
of the action is
\begin{displaymath}
\overline{\phi}_{is} \{ \sigma, \overline{\sigma} \}^i_j \phi^{js}
\: + \: \sqrt{2} \overline{\psi}_{+ is} \overline{\sigma}^i_j
\psi_-^{js} \: + \: {\it c.c.}
\end{displaymath}
At low energies the field $\phi^{is}$ have a vacuum expectation value,
and plugging that in and computing equations of motion yields
\begin{displaymath}
\sigma^i_j \: \propto \: \frac{1}{r} \sum_s \overline{\psi}_{+ j s}
\psi_-^{i s}
\end{displaymath}
The interpretation of $\sigma$ in the low-energy theory follows from
the fact that the $\psi$'s are tangent to the Grassmannian in that theory.

Now, using that interpretation of $\sigma$, let us give a very brief and
heuristic outline of how to apply the methods of \cite{mp} to the
present case.  The superpotential in the
one-loop effective action for the Grassmannian
was calculated in \cite{hori1}[equation (2.16)] along the Coulomb branch,
and has the form
\begin{displaymath}
W \: = \: -t \sum_{a=1}^k \Sigma_a \: - \:
\sum_{a=1}^k N \Sigma_a \left( \log \Sigma_a \: - \: 1 \right).
\end{displaymath}
The fields $\Sigma_a$ are superfields whose lowest components are 
eigenvalues of $\sigma$.  
The vacua have the form $\Sigma_a^N = \exp(-t)$ for
each $a$.
Since $\sigma$ is the curvature of the universal subbundle $S$,
the $\Sigma_a$ should be interpreted as Chern roots of the universal
subbundle.  Naively,  the statement above looks as if it might be
interpreted as a quantum correction to $c_N(S)$, but no such Chern class
exists for $S$.  Instead, using the fact that the Chern classes of
the universal quotient bundle $Q$ are determined by those of $S$,
a more sensible interpretation would be as a quantum correction to
$c_k(S)c_{N-k}(Q)$, which is the form of the quantum cohomology of the
Grassmannian.

Again, this is not meant by any stretch to be a thorough derivation,
but rather is only intended to be a brief heuristic argument linking
the quantum cohomology of the Grassmannian and the one-loop effective
action, in the spirit of \cite{mp}.

A similar heuristic argument can be applied in the case of more
general flag manifolds.  Here, the interpretation of the $\sigma$ fields
is more interesting.  Let us apply the same arguments as above
to a $\sigma$ field coupling to a $U(k_i)$ factor in the gauge group
of a gauged linear sigma model describing $F(k_1, \cdots, k_n, N)$.
Relevant terms in the action are
\begin{displaymath}
- \overline{\phi}_{is} \{ \sigma, \overline{\sigma} \}^i_j \phi^{js}
\: + \: \overline{\tilde{\phi}}_{ia} 
\{ \sigma, \overline{\sigma} \}^i_j \tilde{\phi}^{ja}
\: - \: \sqrt{2} \overline{\psi}_{+ is} \overline{\sigma}^i_j
\psi_-^{js} \: + \: {\it c.c.}
\: + \: \sqrt{2} \overline{\tilde{\psi}}_{+ ia} \overline{\sigma}^i_j
\tilde{\psi}_-^{ja} \: + \: {\it c.c.}
\end{displaymath}
where $\phi^{i s}$, $\psi^{is}$ are bifundamentals in the
$({\bf k_{i-1}}, {\bf \overline{k_i}})$ representation of
$U(k_{i-1}) \times U(k_i)$ and
$\tilde{\phi}^{ia}$, $\tilde{\psi}^{ia}$ are bifundamentals in the
$({\bf k_i},{\bf \overline{k_{i+1}}})$ representation of
$U(k_i) \times U(k_{i+1})$.
Plugging in vacuum expectation values for the $\phi$, $\tilde{\phi}$,
and applying D-terms, we find that
we find that in the low-energy theory the equations of motion for
$\sigma$ have the form 
\begin{equation}   \label{sigmaident}
\sigma^i_j \: \propto \: - \frac{1}{r_i} \sum_s \overline{\psi}_{+ j s}
\psi_-^{i s} \: + \: \frac{1}{r_i} \sum_a \overline{\tilde{\psi}}_{+ j a}
\tilde{\psi}_-^{i a}
\end{equation}
Instead of being the curvature of one universal subbundle $S_i$,
here the field $\sigma$ should now have the interpretation of being
the difference in curvatures between two subbundles.  Indeed,
if we proceed as in \cite{witver} and interpret
the $\overline{\psi}_{+ j s} \psi_-^{is}$ term as the curvature of
$S_{i-1}$, and the $\overline{\tilde{\psi}}_{+ j a} \tilde{\psi}_-^{ia}$
term as the curvature of $S_i$, then we should interpret $\sigma$
as the curvature of $S_i/S_{i-1}$.

Now, let us compute the one-loop effective action for the flag
manifold $F(k_1,\cdots,k_n,N)$.  As in \cite{hori1}, we shall work
along the Coulomb branch in a region where every $U(k_i)$ gauge
symmetry has been Higgsed down to $U(1)^{k_i}$.
In general terms, from \cite{mp}[equ'n (3.36)], the one-loop effective
superpotential in an abelian gauge theory with charged matter should
have the form
\begin{displaymath}
\tilde{W} \: = \:
\sum_a \Sigma_a\left[ - t_a - \sum_i Q_i^a\left( \log\left(
\sum_b Q_i^b \Sigma_b \right) \: - \: 1 \right) \right]
\end{displaymath}
In the case of the Grassmannian $G(k,N)$, there were $N$ fundamentals
of $U(k)$, which became $N$ selectrons of charge 1 under each of the
$U(1)$'s in $U(1)^k$.
In the case of the flag manifold $F(k_1, \cdots, k_n, N)$,
we have bifundamentals $({\bf k_i}, {\bf \overline{k_{i+1}}})$
which become a matrix of selectrons.  The $ij$th entry in the matrix
${\bf C}^{k_i k_{i+1}}$ has charge $+1$ under the $i$th $U(1)$ in
$U(1)^{k_i}$, and charge $-1$ under the $j$th $U(1)$ in $U(1)^{k_{i+1}}$.
Putting this together, we get that the effective superpotential should
be given by
\begin{displaymath}
\tilde{W} \: = \:
\sum_{i=1}^n \sum_{a=1}^{k_i} \Sigma_{ia} \left[
-t_i \: - \: \sum_{s=1}^{k_{i+1}} \left( \log( \Sigma_{ia} \: - \:
\Sigma_{(i+1)s} ) \: - \: 1 \right)
\: + \:
\sum_{s=1}^{k_{i-1}}\left( \log( \Sigma_{(i-1)s} \: - \:
\Sigma_{ia}) \: - \: 1 \right) \right]
\end{displaymath}
with obvious modifications at the endpoints.

For example, in the special case $F(k_1,k_2,N)$, the effective superpotential
is given by
\begin{eqnarray*}
\tilde{W} & = & \sum_{a=1}^{k_1} \Sigma_{1a} \left[ - t_1 \: - \:
\sum_{s=1}^{k_2} \left( \log( \Sigma_{1a} \: - \: \Sigma_{2s} ) \: - \: 1
\right) \right] \\
& & + \: \sum_{a=1}^{k_2} \Sigma_{2a}\left[ -t_2 \: - \: 
\sum_{s=1}^N\left( \log(\Sigma_{2a}) \: - \: 1 \right)
\: + \:
\sum_{s=1}^{k_1}\left[ \log(\Sigma_{1s} \: - \Sigma_{2a}) \: - \: 1
\right) \right]
\end{eqnarray*}
The equations of motion for $\Sigma_{1a}$ are given by
\begin{equation}    \label{sigma1}
\prod_{s=1}^{k_2} \left( \Sigma_{1a} \: - \: \Sigma_{2s} \right)
\: = \: \exp(-t_1) \: \equiv \: q_1
\end{equation}
The equations of motion for $\Sigma_{2a}$ are given by
\begin{eqnarray*}
\Sigma_{2a}^N & = & \exp(-t_2) \prod_{s=1}^{k_1} \left(
\Sigma_{1s} \: - \: \Sigma_{2a} \right) \\
& = & q_2 \prod_{s=1}^{k_1} \left(
\Sigma_{1s} \: - \: \Sigma_{2a} \right)
\end{eqnarray*}
where $q_i \equiv \exp(-t_i)$.

Now, let us compare to the quantum cohomology ring relations for
$F(k_1,k_2,N)$ computed earlier.  We found, for example,
\begin{eqnarray*}
\lefteqn{ x^{(1)}_{k_1} x^{(2)}_{k_2-k_1} \: + \:
x^{(1)}_{k_1-1} x^{(2)}_{k_2-k_1} x^{(3)}_1 \: + \:
x^{(1)}_{k_1} x^{(2)}_{k_2-k_1-1} x^{(3)}_1 \: + \: \cdots
} \\
& & \: + \: 
(-)^{k_2-k_1} q_1 \: + \: (-)^{N-k_2-1} q_2 x^{(1)}_{k_1+k_2-N} \: = \: 0
\end{eqnarray*}
and
\begin{eqnarray*}
\lefteqn{ x^{(2)}_{k_2-k_1} x^{(3)}_{N-k_2} \: + \:
x^{(1)}_1 x^{(2)}_{k_2-k_1-1} x^{(3)}_{N-k_2} \: + \:
x^{(1)}_1 x^{(2)}_{k_2-k_1} x^{(3)}_{N-k_2-1} \: + \: \cdots } \\
& & \: + \:
(-)^{N-k_2-1} q_2 \: + \: (-)^{k_2-k_1} q_1 x^{(3)}_{N-k_2-k_1}
\:  = \: 0
\end{eqnarray*}
where $x^{(i)}_j = c_j(S_i/S_{i-1})$.  The non-quantum-corrected
ring relations allow us to solve for the $x^{(3)}_j$ in terms of the
$x^{(1)}_j$ and $x^{(2)}_j$.  Note that in these polynomial
equations, $q_1$ appears as if it
were a cohomology class of degree $k_2$, and $q_2$ appears as if it were
a cohomology class of degree $N-k_1$.

Now, from equation~(\ref{sigmaident}), we identify $\Sigma_{1a}$ with the
Chern roots of $S_1$ and $\Sigma_{2a}$ with the Chern roots of
$S_2/S_1$, {\it i.e.} the $x^{(1)}_j$ are built from the
$\Sigma_{1a}$ and the $x^{(2)}_j$ are built from the $\Sigma_{2a}$.
In both cases, we find that the $q_i$ are related to polynomials in
the Chern classes -- obtained by expanding out the products
$\prod( \Sigma_{1a} - \Sigma_{2s})$ in the equations of motion for
$\Sigma_{1a}$ and $\Sigma_{2a}$.
Furthermore, and more importantly, note that those same equations of
motion are consistent with $q_1$ behaving as a class of degree $k_2$
and $q_2$ behaving as a class of degree $N-k_1$.

As our purpose in this section was merely to
heuristically outline how the quantum cohomology ring arises in
nonabelian gauged linear sigma models, we shall not pursue this
direction further.

\subsection{Linear sigma model moduli spaces}  \label{qclsm}

In this subsection we shall discuss linear sigma model moduli spaces
for Grassmannians and flag manifolds, which are relevant to
some computations of quantum cohomology rings.
Such moduli spaces are well-understood physically for 
abelian gauged linear sigma models, but to our knowledge have not
been worked out for nonabelian gauged linear sigma models,
where the story is necessarily more complicated.
There is a natural mathematical proposal for such moduli spaces --
namely that they should be Quot schemes and hyperquot schemes --
and we shall see explicitly in examples that the physical
moduli spaces really are precisely the appropriate Quot and
hyperquot schemes, as one would have naively guessed from mathematics,
and not some other related (possibly birational) spaces instead.

Before doing so, however, let us first review some basic facts
regarding honest maps 
\begin{displaymath}
{\bf P}^1 \: \longrightarrow \: F(k_1, \cdots, k_n, N)
\end{displaymath}
Such a map corresponds to a flag of bundles over ${\bf P}^1$:
\begin{displaymath}
{\cal E}_1 \: \hookrightarrow \: {\cal E}_2 \: \hookrightarrow \: \cdots
\: \hookrightarrow \: {\cal E}_n \: \hookrightarrow \:
{\cal O}^{\oplus N}_{ {\bf P}^1 }
\end{displaymath}
where $\mbox{rank } {\cal E}_i = k_i$.
Put simply, this is because a map from ${\bf P}^1$ into the flag
manifold is given by specifying, for each point on ${\bf P}^1$,
a flag, which defines a point on the flag manifold.
Those flags fit together to form a flag of bundles on ${\bf P}^1$.
The multidegree of the map is defined by the degrees of the 
bundles ${\cal E}_i$ above.

\subsubsection{Grassmannians}

What is the linear sigma model moduli space for target $G(k,N)$?

Recall the linear sigma model describes a GIT quotient of
${\bf C}^{kN}$ by $GL(k)$.  The $kN$ chiral superfields $\phi_{is}$
are naturally $N$ copies of the fundamental representation of
$GL(k)$.  The zero modes of the $\phi_{is}$
are given as elements of
\begin{displaymath}
H^0\left( {\bf P}^1, \phi^* ( S^{\vee} \otimes V) \right)
\end{displaymath}

In a nonlinear sigma model on a Grassmannian,
to specify the degree of the map we would need merely specify
an integer, ultimately because $H^2(G(k,N), {\bf Z}) = {\bf Z}$.
In a linear sigma model, however, we must specify a partition
of $d$ into $k$ integers $a_i$.  These integers specify the splitting
type on the ${\bf P}^1$ worldsheet:
\begin{displaymath}
\phi^* S^{\vee} \: = \: \oplus_i {\cal O}(a_i)
\end{displaymath}
where
\begin{displaymath}
\sum_i a_i \: = \: d
\end{displaymath}
In principle, all splitting types should contribute physically,
but the gauged linear sigma model appears to provide a stratification
of the full moduli space.

From Hirzebruch-Riemann-Roch,
\begin{eqnarray*}
\chi\left( S^{\vee} \otimes V \right) & = &
c_1\left( S^{\vee} \otimes V \right) \: + \: kN(1-g) \\
& = & Nd \: + \: Nk \\
& = & N(k+d) 
\end{eqnarray*}

We can expand out the zero modes of each chiral superfield in the usual
fashion.  Note that the original $U(k)$ symmetry is broken by the choice
of $a_i$ to some subset.  Only when all the $a_i$ are the same is the
complete $U(k)$ preserved.

Also note that we can recover an ordinary projective space
in the case that $k=1$, in which case the subtlety above does not
arise.

We shall consider several examples next, and study the various strata
appearing in the gauged linear sigma model moduli space.
Then, after examining several examples, we will discuss how the
complete moduli spaces can be understood, as Quot schemes and
hyperquot schemes.

\subsubsection{Balanced strata example}   \label{firstex}

First, consider degree $km$ maps from ${\bf P}^1$ into $G(k,N)$, so that $km=d$,
and use a partition $a_i=m$ for all $i$.  In this case, the entire
$U(k)$ symmetry will be preserved.
Each chiral superfield can be expanded
\begin{displaymath}
\phi_{is} \: = \: \phi_{is}^0 u^m \: + \: \phi_{is}^1 u^{m-1}v \: + \:
\cdots \: \phi_{is}^{m} v^m
\end{displaymath}
where $u$, $v$ are homogeneous coordinates on the worldsheet.
The coefficients $\phi_{is}^a$ transform in the fundamental representation
of $U(k)$, which suggests that the linear sigma model moduli 
space is given by
\begin{displaymath}
{\bf C}^{(m+1)kN} // GL(k) 
\end{displaymath}
One example of such a quotient, depending upon the excluded set,
is the Grassmannian $G(k,N(m+1))$, but we shall see that this stratum
is not precisely a Grassmannian.

To completely determine this stratum, we need to specify the excluded set.
In order for the LSM moduli space to be given by the Grassmannian
$G(k,N(m+1))$, we would need that for generic $u$, $v$, the
$k$ vectors $(\phi_{i})^a_s = \phi^a_{is}$ in ${\bf C}^{N(m+1)}$ are
orthonormal.  Instead we have a slightly different condition.
The D-terms in the original LSM say that the $\phi_{is}$ should
form $k$ orthonormal vectors in ${\bf C}^N$.
Applying this constraint guarantees that for generic points
on the moduli space, we will be describing a flag of bundles on
${\bf P}^1$, which as discussed earlier describes a map into
a flag manifold.
In terms of zero modes, this implies that for generic $u$, $v$,
\begin{eqnarray*}
\lefteqn{ \sum_s \left( \phi^0_{is} u^m \: + \:
\phi^1_{is} u^{m-1} v \: + \: \cdots \right)
\left( \phi^0_{js} u^m \: + \: \phi^1_{js} u^{m-1} v \: + \: \cdots \right) }\\
& = & \sum_s \phi^0_{is} \phi^0_{js} u^{2m} \: + \:
\sum_s \left( \phi^0_{is} \phi^1_{js} \: + \: \phi^1_{is} \phi^0_{is}
\right) u^{2m-1} v \: + \: \cdots \\
& = & \delta_{ij} r
\end{eqnarray*}
More compactly, we can express this condition by saying that over
every point on ${\bf P}^1$, the $k \times N$ matrix $\phi_{is}$
must be of rank $k$.  To be in the Grassmannian $G(k,N(m+1))$,
by contrast, a weaker condition
need be satisfied, namely that the $k$ vectors in ${\bf C}^{N(m+1)}$
with coordinates $(\phi_i)^a_s = \phi^a_{is}$, must be linearly
independent.
To see that our condition is stronger, consider the
following example.  Take any $k$ and $N$, $m_k-2$.  If the $\phi^a_{is}$ are
defined by, for $s=1$, a degree $m$ polynomial with $k$ different
terms and, for $s>1$, identically zero degree $m$ polynomials,
the result is a perfectly good point of the Grassmannian but which does
not satisfy our stronger condition, as the matrix $\phi_{is}$ is
of rank $1$ not rank $k$.

So, briefly, this component of the moduli space has the form
$U // GL(k)$ for $U$ an open subset of ${\bf C}^{kN(m+1)}$,
and is a proper subset of the Grassmannian $G(k,N(m+1))$.

In the special case that $k=1$, this simplifies slightly, and the
condition to be on the LSM moduli space becomes identical to that for
the Grassmannian, so the moduli space becomes 
$G(1,N(m+1)) = {\bf P}^{N(m+1)-1}$, reproducing the standard result for
linear sigma model moduli spaces of maps ${\bf P}^1 \rightarrow
{\bf P}^{N-1}$.

\subsubsection{Degree 1 maps example}   \label{secondex}

Next, let us consider degree one maps from ${\bf P}^1$ into $G(2,N)$.
We can partition as $(1,0)$ or $(0,1)$.  Physics only sees unordered
partitions, so, there is a single stratum.
Here we run into a subtlety:  since the line bundles in the partition
are no longer symmetric, we can not quotient precisely by
$GL(2)$.

One naive guess would be to restrict to a subgroup of $GL(2)$ that
preserves the stratification, such as $({\bf C}^{\times})^2$.
One would get that the pieces of the moduli space look like
\begin{displaymath}
\left( {\bf C}^N // GL(1) \right) \times
\left( {\bf C}^{2N} // GL(1) \right)
\: = \: {\bf P}^{N-1} \times {\bf P}^{2N-1}
\end{displaymath}

However, that naive guess is not quite right.
Not only is the truncation of $GL(2)$ to $({\bf C}^{\times})^2$ rather 
unnatural, but it also gives pieces of the
wrong dimension.
These moduli spaces should be some sort of compactification of spaces
of maps into Grassmannians, and the space of maps of degree $d$
from ${\bf P}^1$ into the Grassmannian $G(k,N)$ has \cite{stromme}[theorem 2.1]
dimension
$k(N-k) + dN$.  Here, we are considering degree 1 maps from
${\bf P}^1$ into $G(2,N)$, and so our space should have dimension
$2(N-2) + N = 3N -4$.  Instead, the naive guess above has dimension
$3N-2$, which is too big -- by exactly the same amount as the dimension
of that part of the group $GL(2)$ that we truncated.

The correct computation is somewhat more complicated.
The correct symmetry group is the group of global automorphisms
of the vector bundle ${\cal O} \oplus {\cal O}(1)$.
This is not a subgroup of $GL(2)$.
Suppose more generally we were describing
degree $p+q$ maps into $G(2,N)$, and considering a stratum 
defined by the splitting ${\cal O}(p) \oplus {\cal O}(q)$.
Infinitesimal automorphisms will be defined by a matrix whose entries are of the
form
\begin{displaymath}
\left[ \begin{array}{cc}
H^0\left( {\cal O}(p)^{\vee} \otimes {\cal O}(p) \right) = 
H^0\left({\cal O}\right) &
H^0\left( {\cal O}(p)^{\vee} \otimes {\cal O}(q) \right) = 
H^0\left( {\cal O}(q-p) \right) \\
H^0\left( {\cal O}(q)^{\vee} \otimes {\cal O}(p) \right) = 
H^0\left( {\cal O}(p-q) \right) &
H^0\left( {\cal O}(q)^{\vee} \otimes {\cal O}(q) \right) = 
H^0\left( {\cal O} \right)
\end{array}
\right]
\end{displaymath}

In the present example, corresponding to the splitting
${\cal O} \oplus {\cal O}(1)$, this reduces to
\begin{displaymath}
\left[ \begin{array}{cc}
H^0({\cal O}) & H^0({\cal O}(-1)) = 0 \\
H^0({\cal O}(1)) = {\bf C}^2 & H^0({\cal O})
\end{array} \right]
\end{displaymath}
which is the Lie algebra of some non-reductive group that is
an {\it extension} of $({\bf C}^{\times})^2$.
So we should not be quotienting $({\bf C}^N \times {\bf C}^{2N})$
by $({\bf C}^{\times})^2$, but rather
by a group $G$ which has dimension two larger.
As a result, the correct moduli space in this example should have dimension
$3N -4$ instead of $3N-2$, which matches expectations.

Note that the off-diagonal ${\bf C}^2$ acts on the space
${\bf C}^N \times {\bf C}^{2N}$, {\it i.e.}, $1 \times N$ and $2 \times N$
matrices, by leaving the $1 \times N$ matrices invariant,
and subtracts from the $2\times N$ matrices another $2 \times N$ matrix
formed by tensoring a two-vector of ${\bf C}^2$ with the $1 \times N$
matrix.

As a quick check, consider the case of a splitting ${\cal O}(p) \oplus
{\cal O}(q)$.  In this case, one would expect that one should quotient
by $GL(2)$, as we did previously.
From the general story above, the Lie algebra of the correct group has the
form
\begin{displaymath}
\left[ \begin{array}{cc}
H^0({\cal O}) = {\bf C} & H^0({\cal O}(p-p)) = H^0({\cal O}) = {\bf C} \\
H^0({\cal O}(p-p)) = H^0({\cal O}) = {\bf C} & H^0({\cal O}) = {\bf C}
\end{array} \right]
\end{displaymath}
{\it i.e.} $2\times 2$ matrices, exactly matching the Lie algebra of
$GL(2)$.

We can also use the quantum cohomology of the Grassmannian
described earlier to get a quick consistency check of the dimension
of this moduli space.
Recall that in the Grassmannian $G(k,N)$, the quantum cohomology
ring is defined by
\begin{displaymath}
c_k(S^{\vee}) c_{N-k}(Q^{\vee}) \: = \:
q
\end{displaymath}
Roughly, this OPE is realized by a correlation function of the
form 
\begin{displaymath}
< (\overline{\psi} \psi )^{k(N-k)} (\overline{\psi} \psi)^N > \: = \: q
\end{displaymath}
where the $\overline{\psi} \psi$'s correspond to two-forms on the moduli
space, and the correlation function will be nonzero for maps of
fixed degree when the sum\footnote{In general, this statement would be
modified by Euler classes of obstruction bundles, but in this simple
case there is no obstruction bundle, so this reduces to
dimension counting.} of the degrees equals the dimension of the
moduli space.  Since the moduli space here has dimension
$3N-4$, which is the same as $k(N-k) + N$ for $k=2$, we see that
the dimension of the moduli space we just computed is consistent
with the quantum cohomology ring.

\subsubsection{Degree two maps example}   \label{thirdex}

As another example, consider degree two maps from ${\bf P}^1$
into $G(2,N)$.
Here, the relevant partitions are
$(2,0)$, $(1,1)$, $(0,2)$.  As physics only sees unordered partitions,
there are two strata, corresponding to $(2,0)$ and $(1,1)$.

The partition $(1,1)$ gives rise to a stratum that is a large
open subset of
${\bf C}^{4N}//GL(2) = G(2,2N)$.

The partition $(2,0)$ gives rise to the stratum
\begin{displaymath}
( {\bf C}^{3N} \times {\bf C}^N ) // G
\end{displaymath}
where $G$ is an extension of $({\bf C}^{\times})^2$, as discussed previously.

The partitions fit together in the following form:
in the rank 2 case, if $a<b$, the closure of the partition
corresponding to ${\cal O}(a) \oplus {\cal O}(b)$ contains
the partition for ${\cal O}(a-1) \oplus {\cal O}(b+1)$.
(A quick way to see that this is in the closure is to consider the
example 
\begin{displaymath}
{\cal O} \oplus {\cal O} \oplus {\cal O}(1) \: \longrightarrow \:
{\cal O}(1)
\end{displaymath}
of maps between bundles on ${\bf P}^1$.
Letting homogeneous coordinates on ${\bf P}^1$ be denoted $[x,y]$,
take the map above to be defined by the triple $(x,y,\epsilon)$ for
some constant $\epsilon$.
For $\epsilon \neq 0$, the kernel of the bundle map above is
${\cal O}^2$, but for $\epsilon = 0$, the kernel is
${\cal O}(-1) \oplus {\cal O}(1)$.  Thus, ${\cal O}(-1) \oplus
{\cal O}(1)$ lies in the closure of ${\cal O}^2$.)
Roughly speaking, when the bundle is very `balanced,' in the
obvious sense, it is general, whereas when it is very unbalanced,
it is very special.

As in the last section, we can get a quick consistency check of
the dimension of this moduli space from thinking about the
quantum cohomology ring.  As described earlier, the OPE structure
is defined by the relation
\begin{displaymath}
c_k(S^{\vee}) c_{N-k}(Q^{\vee}) \: = \: q
\end{displaymath}
In the present case, for degree two maps, this OPE structure
is consistent with a correlation function of the general form
\begin{displaymath}
< (\overline{\psi} \psi)^{k(N-k)} (\overline{\psi} \psi)^{2(k + (N-k))} > 
\: = \:
q^2
\end{displaymath}
where we are using the fact that there are no Euler classes of obstruction
bundles in the computation.
In order for the correlation function above to be nonzero in this
instanton sector, the moduli space of maps must have dimension
$k(N-k) + 2N$, and indeed, the generic stratum of our moduli space
has dimension $4N-4 = 2(N-2) + 2N$, as desired for $k=2$.

Note that the nongeneric stratum, corresponding to the partition
$(2,0)$, lies in the closure of the stratum corresponding to 
that for the partition $(1,1)$ and so must have a smaller dimension.
This dimension is seen to be $4N-5$, {\it i.e.} this stratum has
codimension 1 in the full moduli space.  This follows immediately from 
examination of the group $G$, which for this stratum has dimension
5, strictly bigger than the dimension of $GL(2)$.

\subsubsection{Interpretation -- Quot and hyperquot schemes}

In abelian GLSM's, the entire LSM moduli space could be constructed
and analyzed at a single stroke, but we have seen that in nonabelian
GLSM's, we only immediately compute strata of the
full LSM moduli space.  One is naturally led to ask, what is the
complete space of which these are strata?

We propose that the answer is that the 
complete LSM moduli spaces in nonabelian GLSM's
for Grassmannians are spaces known as Quot schemes.

Quot schemes generalize both Grassmannians and Hilbert schemes.
They parametrize quotients of bundles.  In the present case,
a (linear sigma model) moduli space of degree $d$ maps
from ${\bf P}^1$ into $G(k,N)$ is given by a Quot scheme describing
rank $k$ bundles on ${\bf P}^1$ together with an embedding into
${\cal O}^N$, or alternatively quotients of ${\cal O}^N$ bundles.
In the special case of maps ${\bf P}^1 \rightarrow G(1,N) = {\bf P}^{N-1}$,
the Quot scheme is the projective space ${\bf P}^{N(d+1)-1}$,
reproducing the standard result for linear sigma model moduli spaces
of maps ${\bf P}^1 \rightarrow {\bf P}^{N-1}$.   

Quot schemes describing bundles on ${\bf P}^1$ ({\it i.e.} 
maps from ${\bf P}^1$ into a Grassmannian) have a natural stratification,
corresponding to the fact that vector bundles over ${\bf P}^1$
decompose into a sum of line bundles, and there are different ways
to distribute the $c_1$ among the various line bundles.

The strata we have seen in the last few examples,
are precisely strata of corresponding Quot schemes, and the
condition on the exceptional set we described in the first example,
is precisely the right condition to reproduce Quot schemes.
In other words, the last few examples have implicitly described
how Quot schemes arise physically in nonabelian gauged linear
sigma models.

The fact that Quot schemes parametrize 
maps from ${\bf P}^1$ into Grassmannians, and hyperquot schemes
parametrize maps from ${\bf P}^1$ into flag manifolds,
is well-known to
mathematicians, see for example\footnote{This is merely a sample list
of references.  It is not intended to be either complete or comprehensive.}
\cite{bdw,cf1,cf2,cf3,kim2,lchen1,stromme}.
What is new here is the understanding that gauged linear sigma models
for Grassmannians also see precisely Quot's, and not some other,
distinct (possibly birational) spaces instead.

See \cite{huylehn} for more information on Quot schemes. 

We have spent a great deal of time studying maps into Grassmannians,
but there are analogous stories in other nonabelian GLSM's.
For example, maps into partial flag manifolds are described mathematically
by hyperquot schemes, which parametrize flags of bundles.
The physical analysis of the GLSM moduli spaces in GLSM's for
flag manifolds is very similar
to what we have done here  in GLSM's for Grassmannians, and the
conclusion is analogous:  GLSM's for flag manifolds see moduli spaces
of maps given by hyperquot schemes, as one would naively expect.
We will outline the analysis and describe an example next,
to support this claim.

\subsubsection{Flag manifolds}

Let us briefly outline the analysis of linear sigma model moduli spaces
for flag manifolds.

Recall the linear sigma model describes $F(k_1,k_2,\cdots,k_n,N)$
as a GIT quotient
\begin{displaymath}
\left( {\bf C}^{k_1 k_2} \times {\bf C}^{k_2 k_3} \times \cdots \times
{\bf C}^{k_n N} \right) // \left( GL(k_1) \times GL(k_2) \times
\cdots GL(k_n) \right)
\end{displaymath}
Thus, instead of a single vector bundle $\phi^ S$,
we now have a collection of vector bundles $\phi^* S_i$,
of rank $k_i$,
one for each factor in the gauge group of the linear sigma model,
together with inclusion maps
\begin{displaymath}
\phi^* S_1 \: \hookrightarrow \: \phi^* S_2 \: \hookrightarrow \:
\cdots \: \hookrightarrow \: \phi^* S_n \:
\hookrightarrow \: {\cal O}_{ {\bf P}^1 }^N
\end{displaymath}

Because our worldsheet is ${\bf P}^1$, each of these vector bundles
is specified topologically by their first Chern class,
denoted $d_i = c_1(\phi^* S_i^{\vee})$, and each vector bundle will completely
split over ${\bf P}^1$, yielding a stratification of the moduli space,
which we claim can be understood globally as a hyperquot scheme.

To fix conventions, we say that a map from ${\bf P}^1$ into the
flag manifold $F(k_1, \cdots, k_n, N)$ has degree
$(d_1, \cdots, d_n)$ if
$c_1(\phi^* S_i^{\vee}) = d_i$.

In these conventions, the zero modes of the bifundamentals in the
$({\bf k_{i-1}}, {\bf \overline{k_i} })$ representation are holomorphic
sections
\begin{displaymath}
H^0\left( {\bf P}^1, \phi^* S_{i-1}^{\vee} \otimes \phi^* S_{i} \right)
\end{displaymath}
(in conventions where $\phi^* S_{i+1} = {\cal O}^N$).
Also, note that since the $\phi^* S_i$'s are vector bundles,
we do not need to require
\begin{displaymath}
c_1\left( \phi^* S_{i-1}^{\vee} \otimes \phi^* S_{i} \right) \: \geq \: 0
\end{displaymath}
for the moduli space to be nonempty.

\subsubsection{Flag manifold example}

Let us consider the example of maps of degree $(2,1)$
from ${\bf P}^1$ into $F(1,2,N)$.
Such a map is specified by two vector bundles over ${\bf P}^1$
$\phi^* S_1^{\vee}$, $\phi^* S_2^{\vee}$ of $c_1$ equal to $2$, $1$, 
and ranks $1$, $2$, respectively.

This information uniquely specifies $\phi^* S_1^{\vee} = {\cal O}(2)$.
Up to irrelevant ordering, this also uniquely specifies
that $\phi^*S_2^{\vee} = {\cal O}(1) \oplus {\cal O}$.

The zero modes of the bifundamentals are given as follows.
The bifundamentals in the $({\bf 1}, {\bf \overline{2}})$ representation
of $U(1) \times U(2)$ have zero modes counted by
\begin{displaymath}
H^0\left( {\bf P}^1, {\cal O}(2-1) \oplus {\cal O}(2-0) \right) \: = \:
H^0\left( {\bf P}^1, {\cal O}(1) \oplus {\cal O}(2) \right) \: = \:
{\bf C}^2 \times {\bf C}^3 \: = \: {\bf C}^5
\end{displaymath}
The remaining bifundamentals, in the $({\bf 2}, {\bf \overline{N}})$
representation, have zero modes counted by
\begin{displaymath}
H^0\left( {\bf P}^1, ({\cal O}(1) \oplus {\cal O})\otimes {\cal O}^N \right) \: = \:
{\bf C}^{2N} \times {\bf C}^N \: = \: {\bf C}^{3N}
\end{displaymath}
 
The linear sigma model moduli space will then be
$( {\bf C}^5 \times {\bf C}^{3N} ) // G$ for some $G$.
In the present case, $G$ will be a product of two factors,
determined by the $U(1)$ and $U(2)$ factors in the gauge group
of the original linear sigma model.
One of the factors in $G$ will be all of $GL(1)$, 
the automorphisms of ${\cal O}(2)$, which acts 
on the ${\bf C}^5$ in ${\bf C}^5\times {\bf C}^{3N}$ with weight one.
The other factor, determined by the $U(2)$ of the linear sigma model
gauge group, will be a non-reductive group defined as
automorphisms of ${\cal O}(1) \oplus {\cal O}(0)$,
an extension of $GL(1)^2$, as in section~\ref{secondex}.
This group, call it $G'$, acts on both the ${\bf C}^5$
and the ${\bf C}^{3N}$ factors.
So, the linear sigma model moduli space has the form
\begin{displaymath}
\left( {\bf C}^5 \times {\bf C}^{3N} \right) // \left( GL(1) \times G' \right)
\end{displaymath}
Moreover, this space is exactly the pertinent hyperquot scheme,
giving us empirical confirmation of the claim that
linear sigma model moduli spaces for maps from ${\bf P}^1$
into Grassmannians and flag manifolds should be precisely
Quot schemes and hyperquot schemes, respectively.

\section{Conclusions}

In this paper, we have described gauged linear sigma models for
flag manifolds and their properties.  In particular,
we have accomplished three things.
First, we have described the gauged linear sigma models that
describe partial flag manifolds, and outlined aspects relevant to
(0,2) GLSM's, which does not seem to have been in the literature previously.

Second, we have described properties of gauged linear sigma models for
complete intersections in flag manifolds.  After reviewing the
Calabi-Yau conditions for a flag manifold, we have discussed various
examples of phenomena in which different geometric K\"ahler phases of
a given GLSM are not related by birational transformation.
We have discussed both Calabi-Yau and non-Calabi-Yau examples of this
phenomenon, and also proposed a mathematical understanding that
should be the right notion to replace `birational' in this context,
namely, Kuznetsov's homological projective duality.
Although nearly every example discussed in this paper was 
a nonabelian GLSM, we also outline how similar phenomena happen
in abelian GLSM's, which will be studied in greater detail
(together with Kuznetsov's homological projective duality)
in \cite{alltoappear}.

Finally, we have outlined how quantum cohomology of
flag manifolds arises within gauged linear sigma models.
In particular, we describe linear sigma model moduli spaces of
maps into Grassmannians and flag manifolds.
The mathematically natural notion is that those spaces are
Quot and hyperquot schemes, and we describe how to see them
explicitly in the physics of nonabelian GLSM's.

\section{Acknowledgements}

We would like to thank A.~Bertram, T.~Braden, A.~Caldararu,
L.~Chen, I.~Ciocan-Fontanine, J.~Distler, M.~Gross, S.~Hellerman,
S.~Katz, and especially 
A.~Knutson, A.~Kuznetsov, and T.~Pantev for useful
conversations.  This paper began as a project with A.~Knutson,
and later grew in part out of the related work
\cite{alltoappear}, done in collaboration with A.~Caldararu, J.~Distler,
S.~Hellerman, and T.~Pantev.  The work of R.~Donagi was
partially supported by NSF grants
DMS-0612992 and Focused Research Grant DMS-0139799 for
``The geometry of superstrings.''


\begin{thebibliography}{199}

\addcontentsline{toc}{section}{References}


\bibitem{WitPhases} E. Witten, ``Phases of N=2 theories in
two dimensions,''
Nucl Phys. {\bf B403} (1993) 159-222, {\tt hep-th/9301042}.

\bibitem{stxglsm} T. Pantev, E. Sharpe, ``GLSM's for gerbes (and other
toric stacks),'' Adv. Theor. Math. Phys. {\bf 10} (2006) 77-121,
{\tt hep-th/0502053}.

\bibitem{witver} E. Witten, ``The Verlinde algebra and the cohomology
of the Grassmannian,'' {\tt hep-th/9312104}.

\bibitem{hori1} K. Hori, D. Tong, ``Aspects of non-Abelian gauge
dynamics in two-dimensional $N=(2,2)$ theories,''
{\tt hep-th/0609032}.

\bibitem{ps4} S. Hellerman, A. Henriques, T. Pantev, E. Sharpe,
M. Ando, ``Cluster decomposition, T-duality, and gerby CFT's,''
{\tt hep-th/0606034}.

\bibitem{kuz1} A. Kuznetsov, ``Homological projective duality,''
{\tt math.AG/0507292}.

\bibitem{kuz2} A. Kuznetsov, ``Derived categories of quadric fibrations and
intersections of quadrics,'' {\tt math.AG/0510670}.

\bibitem{kuz3} A. Kuznetsov, ``Homological projective duality for
Grassmannians of lines,'' {\tt math.AG/0610957}.

\bibitem{alltoappear} A. Caldararu, J. Distler, S. Hellerman, T. Pantev,
and E. Sharpe, ``Non-birational twisted derived equivalences in
abelian GLSM's,'' to appear.

\bibitem{manin} Y. Manin, {\it Gauge field theory and complex geometry},
Springer-Verlag, 1988.

\bibitem{hubsch} T. Hubsch, {\it Calabi-Yau manifolds:  a bestiary for
physicists}, World Scientific, 1992.

\bibitem{ps1} T. Pantev, E. Sharpe, ``Notes on gauging noneffective
group actions,'' {\tt hep-th/0502027}.

\bibitem{ps2} T. Pantev, E. Sharpe, ``String compactifications on
Calabi-Yau stacks,'' Nucl. Phys. {\bf B733} (2006) 233-296,
{\tt hep-th/0502044}.

\bibitem{ps5} E. Sharpe, ``Derived categories and stacks in
physics,'' {\tt hep-th/0608056}.

\bibitem{seibergdual} N. Seiberg, ``Electric-magnetic duality in
supersymmetric non-Abelian gauge theories,''
{\tt hep-th/9411149}.

\bibitem{hp} W. V. D. Hodge, D. Pedoe, {\it Methods of Algebraic Geometry},
volume 1, Cambridge University Press, 1947.

\bibitem{botttu} R. Bott, L. Tu, {\it Differential forms in algebraic
topology}, Springer-Verlag, 1982.

\bibitem{klebstrass} I. Klebanov, M. Strassler, ``Supergravity and a
confining gauge theory:  duality cascades and chiral-symmetry-breaking
resolution of naked singularities,'' {\tt hep-th/0007191}.

\bibitem{strasscas} M. Strassler, ``The duality cascade,''
{\tt hep-th/0505153}.

\bibitem{cortireid} A. Corti, M. Reid, ``Weighted Grassmannians,''
{\tt math.AG/0206011}.

\bibitem{distrev} J. Distler, ``Notes on (0,2) superconformal field theories,''
{\tt hep-th/9502012}.

\bibitem{batyrev2} V. Batyrev, I. Ciocan-Fontanine, B. Kim, D. van Straten,
``Mirror symmetry and toric degenerations of partial flag manifolds,''
{\tt math.AG/9803108}.

\bibitem{batyrev1} V. Batyrev, I. Ciocan-Fontanine, B. Kim, D. van Straten,
``Conifold transitions and mirror symmetry for Calabi-Yau complete
intersections in Grassmannians,'' Nucl. Phys. {\bf B514} (1998) 640-666,
{\tt alg-geom/9710022}.

\bibitem{andreilev} L. Borisov, A. Caldararu, ``The Pfaffian-Grassmannian
derived equivalence,'' {\tt math.AG/0608404}.

\bibitem{harristu} J. Harris, L. Tu, ``On symmetric and skew-symmetric
determinantal varieties,'' Topology {\bf 23} (1984) 71-84.

\bibitem{andreithesis} A. Caldararu, ``Derived categories of twisted
sheaves on elliptic threefolds,''
{\tt math.AG/0012083}.

\bibitem{andreipriv} A. Caldararu, private communication.

\bibitem{tpantevpriv} T. Pantev, private communication.

\bibitem{orlov1} D. Orlov, ``Derived categories of coherent sheaves and
triangulated categories of singularities,''
{\tt math.AG/0503632}.

\bibitem{medercat} E. Sharpe, ``D-branes, derived categories, and
Grothendieck groups,'' Nucl. Phys. {\bf B561} (1999) 433-450, 
{\tt hep-th/9902116}.

\bibitem{mikedc} M. Douglas, ``D-branes, categories, and ${\cal N}=1$
supersymmetry,'' J. Math. Phys. {\bf 42} (2001) 004, {\tt hep-th/0104147}.

\bibitem{beauvilledonagi} A. Beauville, R. Donagi,
``La variet\'e des droits d'une hypersurface cubique de dimension 4,''
C.R. Acad. Sci. Paris. Ser. I Math. 1985.

\bibitem{kuzpriv310} A. Kuznetsov, private communication,
March 10, 2007.

\bibitem{kuzpriv0403} A. Kuznetsov, private communication, April 3, 2007.

\bibitem{grosspriv} M. Gross, private communication, September 27, 2006.

\bibitem{gh} P. Griffiths and J. Harris, {\it Principles of Algebraic
Geometry}, John Wiley \& Sons, New York, 1978.

\bibitem{mukai1} S. Mukai, ``Moduli of vector bundles on K3 surfaces,
and symplectic manifolds,'' Sugaku Expositions {\bf 1} (1988) 139-174.

\bibitem{bdw} A. Bertram, G. Daskalopoulos, and R. Wentworth,
``Gromov invariants for holomorphic maps from Riemann surfaces
to Grassmannians,'' J. Amer. Math. Soc. {\bf 9} (1996) 529-571.   


\bibitem{sadov1} A. Ashtashkevich, V. Sadov, ``Quantum cohomology of
partial flag manifolds $F_{n_1 \cdots n_k}$,''
Comm. Math. Phys. {\bf 170} (1995) 503-528.  

\bibitem{cf1} I. Ciocan-Fontanine, ``On quantum cohomology rings of
partial flag varieties,'' Duke Math. J. {\bf 98} (1999) 485-524.  

\bibitem{cf2} I. Ciocan-Fontanine, ``The quantum cohomology ring
of flag varieties,'' Trans. Amer. Math. Soc. {\bf 351} (1999)
2695-2729.    

\bibitem{cf3} I. Ciocan-Fontanine, ``Quantum cohomology of flag
varieties,''
Int. Math. Res. Notices 1995, no. 6, 263-277.

\bibitem{kim1} B. Kim, ``Quantum cohomology of partial flag manifolds
and a residue formula for their intersection pairings,''
Int. Math. Res. Notices 1995, no. 1, 1-16.   

\bibitem{kim2} B. Kim, ``Quot schemes for flags and Gromov invariants for
flag varieties,'' {\tt alg-geom/9512003}.

\bibitem{bck0} A. Bertram, I. Ciocan-Fontanine, B. Kim,
``Two proofs of a conjecture of Hori and Vafa,''
Duke Math. J. {\bf 126} (2005) 101-136,
{\tt math.AG/0304403}.

\bibitem{bck} A. Bertram, I. Ciocan-Fontanine, B. Kim,
``Gromov-Witten invariants for abelian and nonabelian quotients,''
{\tt math.AG/0407254}.

\bibitem{lchen1} L. Chen, ``Poincare polynomials of hyperquot schemes,''
{\tt math.AG/0003077}.  

\bibitem{stromme} S. A. Stromme, ``On parametrized rational curves in
Grassmann varieties,'' pp. 251-272 in {\it Space Curves}, Lect. notes in
math. 1266, ed. F. Ghione, C. Peskine, E. Sernesi, Springer-Verlag, 1987.

\bibitem{llly} B. Lian, C.-H. Liu, K. Liu, S.-T. Yau,
``The $S^1$ fixed points in Quot schemes and mirror principle
computations,'' {\tt math/0111256}.

\bibitem{mp} D. Morrison, R. Plesser, ``Summing the instantons:
quantum cohomology and mirror symmetry in toric varieties,''
Nucl. Phys. {\bf B440} (1995) 279-354,
{\tt hep-th/9412236}.


\bibitem{huylehn} D. Huybrechts, M. Lehn, {\it The geometry of moduli spaces
of sheaves}, Vieweg, 1997.










\end{thebibliography}
\end{document}